\documentclass[prd,nofootinbib,preprint,superscriptaddress]{revtex4}
\DeclareUnicodeCharacter{2212}{-}
\pdfoutput=1
\usepackage[T1]{fontenc}
\usepackage{amsmath,amssymb, mathtools}
\usepackage{epsfig}
\usepackage{subfigure}
\usepackage{float}
\usepackage{graphicx}
\usepackage{xcolor} 
\usepackage{slashed}
\usepackage{multirow}
\usepackage[colorlinks,citecolor=blue]{hyperref} 
\usepackage{pdfpages}
\usepackage{comment}

\begin{document}
\newcommand\myeq{\mathrel{\overset{\makebox[0pt]{\mbox{\normalfont\tiny\sffamily vev}}}{=}}}
\title{Electroweak Phase Transition in Two Scalar Singlet Model with pNGB Dark Matter}
\author{Dilip Kumar Ghosh}
\email{dilipghoshjal@gmail.com}
\affiliation{School of Physical Sciences, \\Indian Association for Cultivation of Science, Jadavpur, Kolkata 700032, India}

\author{Koustav Mukherjee}
\email{koustav.physics1995@gmail.com}
\affiliation{School of Physical Sciences, \\Indian Association for Cultivation of Science, Jadavpur, Kolkata 700032, India}

\author{Shourya Mukherjee}
\email{mshourya@umd.edu}
\affiliation{Department of Physics and Astronomy, University of British Columbia, 6224 Agricultural Road, Vancouver, B.C. V6T 1Z1, Canada}
\affiliation{Maryland Center for Fundamental Physics, Department of Physics, University of Maryland, College Park, MD 20742, U.S.A.}

\begin{abstract}
We investigate the dynamics of the electroweak phase transition within an extended Standard Model framework that includes one real scalar $(\Phi)$ and one complex scalar $(S)$, both of which are SM gauge singlets. The global $U(1)$ symmetry is softly broken to a $\mathcal{Z}_3$ symmetry by the $S^3$ term in the scalar potential. After this $U(1)$ symmetry breaking, the imaginary component of the complex scalar $(S)$ acts as a pseudo-Nambu-Goldstone boson (pNGB) dark matter candidate, naturally stabilized by the $\mathcal{Z}_2$ symmetry of the scenario. Specially, the spontaneous breaking of the global $U(1)$ symmetry to a discrete $\mathcal{Z}_3$ subgroup can introduce effective cubic terms in the scalar potential, which facilitates a strong first-order phase transition. We analyze both single-step and multi-step first-order phase transitions, identifying the parameter space that satisfies the dark matter relic density constraints, complies with all relevant experimental constraints, and exhibits a strong first-order electroweak phase transition. The interplay of these criteria significantly restricts the model parameter space, often leading to an underabundant relic density. Moreover, we delve into the gravitational wave signatures associated with this framework, offering valuable insights that complement traditional dark matter direct and indirect detection methods.

\end{abstract}

\maketitle

\section{Introduction}
Dark matter (DM), a mysterious substance of the universe constituting 26$\%$ of the universe's energy density \cite{Planck:2018vyg}, possesses an elusive nature, revealing its presence primarily through gravitational interactions. The quest to unravel the nature of DM and its fundamental constituents has led to the exploration of diverse theoretical frameworks within particle physics, see \cite{Arcadi:2017kky,Bernal:2017kxu} for the review of such frameworks. Among them, over the past few decades, the framework that has gained immense popularity and extensive attention is the paradigm commonly referred to as the Weakly Interacting Massive Particle (WIMP) model \cite{Kolb:1990vq,Feng:2010gw,Roszkowski:2017nbc,Schumann:2019eaa, Lin:2019uvt}. Despite extensive efforts in recent years, WIMP direct detection (DD) experiments, such as XENON1T \cite{XENON:2018voc}, PandaX-4T \cite{PandaX-4T:2021bab}, LUX-ZEPLIN (LZ) \cite{LZ:2022lsv}, have yet to produce any conclusive evidence of such WIMP like dark matter particle. These negative search results can be explained by suppressing the WIMP-nucleon scattering processes while keeping the DM annihilation rate unsuppressed to satisfy the observed DM relic density. In recent years, pseudo-Nambu-Goldstone boson (pNGB) dark matter models have attracted considerable interest due to their inherent suppression of direct detection signals. This feature was first reported in Ref.\cite{Gross:2017dan}, where the cancellation mechanism in Higgs portal was identified, leading to a suppressed tree level dark matter nucleus scattering cross-section. The minimal set-up of the pNGB DM model includes only a complex scalar singlet $(S)$ in conjunction with the SM Higgs doublet $(H)$. The imaginary component of this newly added scalar acts as the pNGB DM after the symmetry is spontaneously broken and the production of such DM particles solely depends on the Higgs portal interactions.  In such pNGB DM models  \cite{Gross:2017dan,Huitu:2018gbc,Alanne:2018zjm,Azevedo:2018oxv,Karamitros:2019ewv,Cline:2019okt,Arina:2019tib}, the global $U(1)$ symmetry undergoes explicit breaking by the mass of DM resulting to a remnant $\mathcal{Z}_2$ symmetry.  This results in the suppression of direct detection signals at the tree level ($t$-channel process), particularly in the limit of zero momentum transfer \cite{Gross:2017dan}. However, the $s$-channel-dominated dark matter pair annihilation processes enable the correct relic density to be achieved, similar to standard WIMP scenarios. The pNGB DM has gained significant attention recently and has been studied in different contexts such as the UV completion \cite{Abe:2020iph,Okada:2020zxo,Abe:2021byq,Okada:2021qmi}, collider constraints \cite{Huitu:2018gbc,Cline:2019okt,Arina:2019tib}, direct detection from loop correction \cite{Azevedo:2018exj,Ishiwata:2018sdi,Glaus:2020ihj,Alanne:2018zjm,Arina:2019tib,Zeng:2021moz}, indirect searches.

The detection of gravitational waves by the LIGO collaboration \cite{LIGOScientific:2016aoc} has steered a new era of precision cosmology, providing a unique window into the physics of the early universe. Gravitational waves, characterized by their distinct sources and detectability \cite{Kuroda:2015owv}, offer invaluable insights into cosmic phenomena. Among these, cosmological first-order phase transitions stand out as one of the promising sources of stochastic gravitational waves via the nucleation of the bubbles of the broken phase.  The electroweak phase transition in the Standard Model with the observed Higgs boson mass $\sim 125$ GeV, is characterized as a crossover rather than a first-order phase transition \cite{PhysRevLett.77.2887, Kajantie:1996qd} and thus precludes the generation of the stochastic gravitational waves \cite {PhysRevLett.77.2887}. However, various well-motivated extensions to the SM have been proposed that can generate first-order phase transitions and eventually GW signals. Several studies on FOPT and GW has been done in various BSM Models like a single  scalar singlet extension \cite{Hashino:2016xoj}, 2 scalar singlet extension \cite{Shajiee:2018jdq,Ghorbani:2019itr,Ghorbani:2024twk}, 2HDM \cite{Goncalves:2021egx}. Such GW waves are potentially discoverable by future space-based interferometers, such as LISA \cite{LISA:2017pwj, Baker:2019nia}, DECIGO \cite{Seto:2001qf, Kawamura:2006up, Yagi:2011wg, Isoyama:2018rjb}, BBO \cite{Crowder:2005nr, Corbin:2005ny, Harry:2006fi}.

Intriguingly, the interplay between an additional scalar singlet and the Standard Model Higgs doublet can potentially induce two-step phase transitions. However, achieving a strong first-order phase transition (SFOPT) within the confines of a $\mathcal{Z_{\rm 2}}$ symmetric model necessitates a substantial number of additional scalar fields, typically around twelve, as demonstrated in \cite{PhysRevD.78.123528}. Furthermore, investigations into phase transitions in conjunction with pNGB dark matter within models featuring a softly broken global $U(1)$ symmetry by a $\mathcal{Z_{\rm 2}}$ subgroup has revealed that all transitions leading to the true vacuum are of second order \cite{Kannike:2019wsn}. Consequently, such scenarios are incompatible with the production of gravitational waves.

To induce a first-order phase transition, explicit symmetry-breaking terms, such as soft cubic interactions \cite{Kannike:2019mzk} or the most general symmetry-breaking operators \cite{Alanne:2020jwx}, are often introduced at the tree level scalar potential. In particular, the spontaneous breaking of a global $U(1)$ symmetry to a discrete $\mathcal{Z_{\rm 3}}$ subgroup \cite{Kannike:2019mzk} can generate effective cubic terms in the scalar potential, thereby facilitating a strongly first-order phase transition \cite{Witten:1984rs, Hogan:1983ixn, Steinhardt:1981ct}. In such models, both the broken and unbroken phases stabilize the dark matter candidate due to the symmetry $S\rightarrow S^{\dagger}$ present in the scalar potential where $S$ is the complex singlet scalar. The main difference between the $\mathcal{Z_{\rm 2}}$ and $\mathcal{Z_{\rm 3}}$ is the presence of the cubic terms in the latter which respect the $\mathcal{Z_{\rm 3}}$ symmetry of the potential. Apart from suppressing the direct detection cross section \cite{Kannike:2019mzk}, the presence of the cubic terms generates a potential barrier at the tree level \cite{Ghosh:2022fzp} promoting the occurrence of a strong first-order phase transition. The production of gravitational waves (GWs) during such phase transitions has been studied in BSM scenarios with additional scalar fields \cite{Espinosa:2011ax, Alanne:2016wtx, Cheng:2018ajh}. The phenomenological implications of first-order phase transitions and gravitational wave production within a $\mathcal{Z_{\rm 3}}$-symmetric model featuring a complex scalar singlet were reported in \cite{Kang:2017mkl, Kannike:2019mzk,Chiang:2019oms}.

In this work, we propose a thorough investigation of DM phenomenology, electroweak phase transition (EWPT) dynamics, and subsequent GW production within a $\mathcal{Z_{\rm 3}}$-symmetric model that incorporates two scalar singlets (one real, one complex) in conjunction with the Standard Model Higgs doublet. Our extended framework aims to address the challenges encountered in previous studies by potentially enhancing the phase transition strength through the introduction of additional degrees of freedom. The inclusion of a real singlet scalar, invariant under a spontaneously broken $\mathcal{Z_{\rm 2}}$ symmetry, offers a minimal yet effective extension to accommodate both SFOPT and viable DM phenomenology while adhering to current experimental constraints.

 In our analysis, we first scan the multi-dimensional parameter space that satisfies various theoretical and experimental constraints such as vacuum stability, perturbative unitarity, electroweak precision observables (S \& T) parameters, etc. In this framework, we have two real scalars, in addition to the SM Higgs, thus the field configuration space becomes 3-dimensional. This multi-dimensional field configuration makes the shape of the phase transitions more intriguing. We use the publicly available  \textsf{CosmoTransitions}\cite{Wainwright:2011kj}package to generate the results of the phase structures of the scalar fields. We find seven benchmark points (BP1-BP7) that show strong first-order phase transitions. Out of these, three benchmark points (BP1, BP2 \& BP3) show SFOPT along both the SM and BSM Higgs field directions, while for BP4-BP7, we get SFOPT along the BSM Higgs direction only. As mentioned before, the pNGB dark matter is an integral  part of this scenario. Therefore, the next obvious scrutiny would be to check whether the aforementioned seven benchmark points are consistent with the DM relic density, direct and indirect search limits. The seven benchmark  points are consistent with dark matter relic density.  Most of the points show underabundant relic density except for BP3 and BP7 and thus can be a viable dark matter component. However, it turns out that for some of these benchmark points, the dark  matter direct detection cross-section is too large to be satisfied by the current DM direct detection bounds due to the simultaneous requirement of SFOPT and correct relic density which requires high and low values of the scalar quartic couplings respectively. Points where SFOPT is observed and all the latest dark matter experimental results are satisfied, the dark matter relic density is highly underabundant. Further, we show the GW spectrum of those points and whether they would possibly be detected in upcoming future experiments like DECIGO and U-DECIGO.

The paper is organized as follows. The model is introduced in Sec.\ref{model}. The theoretical and experimental constraints are discussed in Sec.\ref{constraints}. Details of FOPT are discussed in Sec.\ref{phase transition} along with the benchmark points of our model. Dark matter phenomenology is discussed in Sec.\ref{dark-matter}. The GW spectrum corresponding to our chosen benchmark points is studied in Sec.\ref{Gravitywave}. Finally, we conclude in Sec.\ref{conclusion}. Additional details of field dependent mass matrix of the scalar fields, co-positivity and perturbative unitarity constraints, expressions of counter terms appearing due to finite temperature corrections, etc. are given in the Appendix.

\section{The Model}
\label{model}
The simplest scenario for the pseudo-Nambu Goldstone (pNGB) dark matter consists of a complex gauge singlet scalar $S$ with softly broken global $U(1)$ symmetry to provide a mass term for the pNGB dark matter \cite{Gross:2017dan}. However, in this work, we extend the minimal framework of pNGB DM by another gauge singlet real scalar $\Phi$. We also introduce $\mathcal{Z}_2\times \mathcal{Z}_3$ discrete symmetry. The $S$ field does not carry any $\mathcal{Z_{\rm 2}}$ charge while under $\mathcal{Z}_{\rm 3}$ it transforms as $S\rightarrow e^{\frac{2\pi i}{3}} S$. In contrast, $\Phi$ has odd charge under $\mathcal{Z_{\rm 2}}$, however remain invariant under $\mathcal{Z}_3$. The most general renormalizable scalar potential containing the SM Higgs doublet $H$, complex scalar $S$ and the real scalar $\Phi$ is given by  

\begin{align}
V_{\rm 0}(H,S,\Phi) =\,& {\mu_{H}^2}\,( H^\dagger H)+{\lambda_{H}}\, 
( H^\dagger H)^2 +{\mu_{\Phi}^2}\,\Phi^2 + \lambda_{\Phi} \,  \Phi^4 + {\mu_{S}^2}\,( S^\dagger S) +\lambda_{S} \, ( S^\dagger S)^2+ \nonumber \\& \lambda_{H S}\,
(H^\dagger H)(S^\dagger S)+\lambda_{S \Phi}(S^\dagger S) \Phi^2+\lambda_{H \Phi}(H^\dagger H)\Phi^2 +\frac{\mu_3}{2}(S^3+S^{\dagger3}). 
\label{eq:potential}
\end{align}
 The $\mu_3$ term in Eq.(\ref{eq:potential}) softly breaks the global $U(1)$ symmetry. In addition, the potential has discrete symmetry $\mathcal{Z}_2$ symmetry, $S\to S^\dagger $, just like the original pNGB dark matter model. This $\mathcal{Z}_2$ symmetry, as imposed, also prevents the odd power terms of $\Phi$ from appearing in the potential.

In the unitary gauge, after electroweak symmetry breaking (EWSB), we parametrize the scalar fields in the following form:
\begin{align}
    H=\begin{bmatrix}
      0 \\
      \frac{v_{h}+h}{\sqrt{2}}
    \end{bmatrix};~~
    S=\frac{v_s+s+i\chi}{\sqrt{2}};~~
    \Phi=\phi+v_{\phi}.
\end{align}
where, $v_h, v_s $ and $v_\phi$ represent vacuum expectation values for the $H, S$ and $\Phi$ fields respectively. Throughout our analysis we assume these vacuum expectation values to be real. The non-zero vev of $S$ field, $\langle S \rangle = v_s\neq 0$, spontaneously breaks the $\mathcal{Z}_3$ symmetry and the pNGB field $\chi$ transforms as $\chi\rightarrow -\chi$ under the $\mathcal{Z}_2$ symmetry and therefore it qualifies as a stable DM candidate even if $\mathcal{Z}_3$ is broken. \footnote{In general, spontaneous breaking of discrete $\mathcal{Z_{\rm 3}}$ symmetry leads to the formation of domain walls. However, this the issue of domain wall creation can be avoided if $U(1)$ symmetry is gauged in a UV complete model \cite{Gross:2017dan,Abe:2022mlc,Dvali:1994wv}.}

The minimisation conditions of the scalar potential in Eq.(\ref{eq:potential}) result into following relations,
\begin{align}
    \mu_H^2 = & - \lambda_H v_h^2  - \frac{\lambda_{HS}}{2}v_s^2-\lambda_{H \Phi } v_{\phi}^2,\\
    \mu_S^2 =& - \lambda_{S \phi} v_{\phi}^2  - \lambda_S v_s^2 - \frac{\lambda_{HS}}{2} v_h^2 - \frac{3 \sqrt{2} \mu_3}{4}v_s,\\
    \mu_{\Phi}^2 =& -2 \lambda_{\Phi} v_{\phi}^2  - \frac{\lambda_{H \Phi}}{2} v_h^2 - \frac{\lambda_{S\Phi}}{2} v_s^2.
\end{align}

In our scenario, with the possibilities of three vevs $v_h,v_s$ and $v_\phi$, we can have the following five types of extrema in the $(v_h, v_s, v_\phi)$ plane :
\begin{enumerate}
    \item Unbroken phase : ${\mathcal O}_S \equiv (0,0,0) $;
    \item Spontaneous breaking of the SM Higgs vev : ${\mathcal O}_H \equiv (v_h, 0,0)$;
    \item Spontaneous breaking of $\mathcal{Z}_3$ symmetry of the $S$ field : ${\mathcal O}_S \equiv (0,v_s,0)$;
    \item Spontaneous breaking of $\mathcal {Z}_2$ symmetry of the real $\Phi$ field : ${\mathcal O}_\Phi \equiv (0,0,v_\phi)$;
    \item Spontaneous breaking of all three symmetries : ${\mathcal O}_{HS\Phi} \equiv (v_h,v_s,v_\phi)$
\end{enumerate}

\noindent We are interested in the symmetry-broken phase of the scalar potential. We start our analysis in ${\mathcal O}_{HS\Phi}$ phase, where the mass squared matrix of the CP-even scalar fields at tree level is given by
\begin{equation}
\centering
\mathcal{M}^2_{\rm scalar} = \begin{bmatrix}
 2 \lambda_h v_h^2 & \lambda_{HS} v_h v_s & 2 \lambda_{H\Phi} v_h v_\phi \\
    \lambda_{HS} v_h v_s  & 2\lambda_S v_S^2+\frac{3\mu_3 v_s}{2\sqrt{2}} & 2 \lambda_{S\Phi} v_s v_\phi \\
    2 \lambda_{H\Phi} v_h v_\phi  & 2 \lambda_{S\Phi} v_s v_\phi  & 8\lambda_\phi v_\phi^2
    \end{bmatrix}\label{eq:masssquareCPeven}
\end{equation}

\noindent We diagonalize the scalar mass squared matrix by the transformation $O^T\mathcal \mathcal{M}_{\rm scalar}^2 O = {\rm diag} (m^2_{H_1}, m^2_{H_2}, m^2_{H_3})$. Here $O$ is a $3\times 3 $ orthogonal matrix comprising of three rotational angles $\alpha_1,\alpha_2$ and $\alpha_3$ which is given by,
\begin{equation}
\centering
O_{\alpha} =   \begin{bmatrix}
\label{matrix2}
    c_{\alpha_1} c_{\alpha_2} & s_{\alpha_1}c_{\alpha_2} & s_{\alpha_2} \\
    -(c_{\alpha_1}s_{\alpha_2}s_{\alpha_3}+s_{\alpha_1}c_{\alpha_3}) & c_{\alpha_1}c_{\alpha_3}-s_{\alpha_1}s_{\alpha_2}s_{\alpha_3} & c_{\alpha_2}s_{\alpha_3} \\
    -c_{\alpha_1}s_{\alpha_2}c_{\alpha_3}+s_{\alpha_1}s_{\alpha_3} & -(c_{\alpha_1}s_{\alpha_3}+s_{\alpha_1}s_{\alpha_2}c_{\alpha_3}) & c_{\alpha_2}c_{\alpha_3}
\end{bmatrix}
\end{equation}

\noindent We use the shorthand notation: $c_{\alpha_{1,2,3}} \equiv \cos \alpha_{1,2,3}$ and $s_{\alpha_{1,2,3}} \equiv \sin \alpha_{1,2,3}$. The mixing angles can be chosen to lie in the range:
\begin{eqnarray}
-\frac{\pi}{2} < \alpha_1, \alpha_2, \alpha_3 < \frac{\pi}{2}
\end{eqnarray}
without loss of generality. The orthogonal matrix $O_\alpha$ can be expressed as a product of three rotation matrices ${\mathcal R}(\alpha_3), {\mathcal R}(\alpha_2), {\mathcal R}(\alpha_1)$, such that $O_\alpha ={\mathcal R}(\alpha_3).{\mathcal R}(\alpha_2).{\mathcal R}(\alpha_1)$
with \begin{equation}\hspace*{-0.3cm} \label{eq:R1R2R3}
{\mathcal R}({\alpha_1}) = \begin{pmatrix} 
c_{\alpha_{1}} & s_{\alpha_{1}} & 0 \\ -s_{\alpha_{1}} & c_{\alpha_{1}} & 0 \\ 0 & 0 & 1  \end{pmatrix},~~
{\mathcal R}({\alpha_2}) = \begin{pmatrix} c_{\alpha_{2}} & 0 & s_{\alpha_{2}} \\ 0 & 1 & 0 \\ -s_{\alpha_{2}} & 0 & c_{\alpha_{2}}  \end{pmatrix},~~ {\mathcal R}({\alpha_3}) = \begin{pmatrix} 1 & 0 & 0 \\  0 & c_{\alpha_{3}} & s_{\alpha_{3}} \\ 0 &   -s_{\alpha_{3}} & c_{\alpha_{3}}\end{pmatrix}.
\end{equation}

\noindent The mixing of three CP even gauge eigenstates $(h,s, \phi)$ will yield to the following three CP even physical eigenstates:
\begin{align}
\begin{bmatrix}
 H_1\\
 H_2\\
 H_3
\end{bmatrix}=O^{T}(\alpha_1,\alpha_2,\alpha_3)
\begin{bmatrix}
 h\\
 s\\
 \phi
\end{bmatrix}
\end{align}
Thus the relation between the physical basis ($H_1,H_2,H_3$) with the unphysical basis ($h,s,\phi$) is given by,
\begin{eqnarray}
    H_1 &=& (c_{\alpha_1} c_{\alpha_2}) h -(c_{\alpha_1}s_{\alpha_2}s_{\alpha_3}+s_{\alpha_1}c_{\alpha_3}) s + (s_{\alpha_1}s_{\alpha_3}-c_{\alpha_1}s_{\alpha_2}c_{\alpha_3}) \phi \\
    H_2 &=& (s_{\alpha_1}c_{\alpha_2}) h + (c_{\alpha_1}c_{\alpha_3}-s_{\alpha_1}s_{\alpha_2}s_{\alpha_3}) s -(c_{\alpha_1}s_{\alpha_3}+s_{\alpha_1}s_{\alpha_2}c_{\alpha_3}) \phi \\
    H_3 &=&  (s_{\alpha_2}) h + (c_{\alpha_2} s_{\alpha_3}) s + (c_{\alpha_2}c_{\alpha_3}) \phi
\end{eqnarray}

\noindent In this analysis, we assume the following mass hierarchy among these CP even scalars :
\begin{eqnarray}
    m^2_{H_3} > m^2_{H_2} > m^2_{H_1}
\end{eqnarray}

\noindent where the lightest CP even scalar $H_1$ is identified to the Standard Model (SM) Higgs boson $(h_{SM})$ having mass $m_{H_1} = 125$ GeV, and $v_h = v_{\rm sm} = 246$ GeV respectively. Throughout the analysis, we ensure that the properties of $H_1$ remain consistent with the experimentally measured values of the Standard Model Higgs boson at the LHC. On the other hand, the corresponding mass of the pseudoscalar (pNGB) $\chi$ is by,
\begin{align}
    m_\chi^2=-9\frac{\mu_3 v_s}{2\sqrt{2}}
    \label{mass:chi}
\end{align}

\noindent Note that, $m^2_\chi$ is proportional to $\mu_3$ indicating that the soft breaking of the $U(1)$ symmetry is solely responsible for giving mass to the pNGB dark matter candidate. In the absence of this soft-breaking term, the massive pNGB would become a massless Nambu-Goldstone boson upon the spontaneous breaking of the global $U(1)$ symmetry.

We express all the parameters of the scalar potential in terms of physical quantities $m^2_{H_i},{(i=1-3)},\alpha_1, \alpha_2,\alpha_3, v_h, v_s, v_\phi $ and the dark matter mass $m^2_\chi$, measured at zero temperature vacuum:
\begin{align}
   & \lambda_{H}  = \frac{1}{2 v_{h}^2}\left[m^2_{H_{1}}c^2_{\alpha_{1}}c^2_{\alpha_{2}}+m^2_{H_{2}}(c_{\alpha_{3}}s_{\alpha_{1}}+c_{\alpha_{1}}s_{\alpha_{2}}s_{\alpha_{3}})^2
   + m^2_{H_{3}}(s_{\alpha_1}s_{\alpha_3}-c_{\alpha_1}s_{\alpha_2}c_{\alpha_3})^2\right],\nonumber\\
  & \lambda_{S} = \frac{1}{2 v_{s}^2}\left[m^2_{H_{1}}s_{\alpha_{1}}^2c_{\alpha_{2}}^2+m^2_{H_{2}}(c_{\alpha_{3}}c_{\alpha_{1}}-s_{\alpha_{1}}s_{\alpha_{2}}s_{\alpha_{3}})^2
   + m^2_{H_{3}}(c_{\alpha_{1}}s_{\alpha_{3}}+c_{\alpha_{3}}s_{\alpha_{1}}s_{\alpha_{2}})^2 -\frac{3}{2\sqrt{2}}\mu_{3}v_{s}\right],\nonumber\\\
  &\lambda_{\Phi} =  \frac{1}{8 v_{\phi}^2}\left[m^2_{H_{1}}s_{\alpha_{2}}^2+m^2_{H_{2}}c_{\alpha_{2}}^2s_{\alpha_{3}}^2
   + m^2_{H_{3}}c_{\alpha_{2}}^2c_{\alpha_{3}}^2 \right],\nonumber\\
 & \lambda_{HS} = \frac{1}{ v_{h}v_{s}}\Big[m^2_{H_{1}}s_{\alpha_{1}}c_{\alpha_{1}}c_{\alpha_{2}}^2-
   m^2_{H_{2}}(c_{\alpha_{3}}s_{\alpha_{1}}+c_{\alpha_{1}}s_{\alpha_{2}}s_{\alpha_{3}})(c_{\alpha_{1}}c_{\alpha_{3}}-s_{\alpha_{1}}s_{\alpha_{2}}s_{\alpha_{3}})
    \nonumber\\
   &~~~~~~~~~~~~~~~~~~~~~~~~~~~~~~~~~~~~ +m^2_{H_{3}}(s_{\alpha_{1}}s_{\alpha_{3}}-c_{\alpha_{1}}s_{\alpha_{2}}c_{\alpha_{3}})(-c_{\alpha_{3}}s_{\alpha_{2}}s_{\alpha_{1}}-c_{\alpha_{1}}s_{\alpha_{3}}) \Big],\nonumber\\
   &\lambda_{S\Phi}= \frac{1}{ 2v_{\phi}v_{s}}\Big[m^2_{H_{1}}s_{\alpha_{1}}s_{\alpha_{2}}c_{\alpha_{2}}+
   m^2_{H_{2}}c_{\alpha_{2}}s_{\alpha_{3}}(c_{\alpha_{1}}c_{\alpha_{3}}-s_{\alpha_{1}}s_{\alpha_{2}}s_{\alpha_{3}})-
   m^2_{H_{3}}(c_{\alpha_{2}}c_{\alpha_{3}})(c_{\alpha_{3}}s_{\alpha_{2}}s_{\alpha_{1}}+c_{\alpha_{1}}s_{\alpha_{3}}) \Big],\nonumber\\
   & \lambda_{H\Phi} = \frac{1}{ 2v_{\phi}v_{h}}\Big[m^2_{H_{1}}c_{\alpha_{1}}c_{\alpha_{2}}s_{\alpha_{2}}-
m^2_{H_{2}}c_{\alpha_{2}}s_{\alpha_{3}}(c_{\alpha_{3}}s_{\alpha_{1}}+c_{\alpha_{1}}s_{\alpha_{2}}s_{\alpha_{3}})+
   m^2_{H_{3}}(c_{\alpha_{2}}c_{\alpha_{3}})(s_{\alpha_{1}}s_{\alpha_{3}}-c_{\alpha_{1}}s_{\alpha_{2}}c_{\alpha_{3}}) \Big],
   \label{qcoupling}\end{align}
 
\section{Theoretical and Experimental Constraints}\label{constraints}
In this section, we scrutinize various theoretical and experimental constraints on the parameter space of our proposed scenario. To achieve this, we scan over the following free model parameters: 
  \begin{eqnarray}
  \label{eqw:int_parameter}
      \{m_{H_2},m_{H_3},m_\chi,\sin\alpha_1,\sin\alpha_2,\sin\alpha_3,v_s,v_\phi\}
  \end{eqnarray} 
within the range as shown in Table \ref{paramscanrange}.

\setlength{\tabcolsep}{12pt}
\begin{table}[]
    \centering
    \begin{tabular}{|c|c|}
        \hline
        Parameters & Scanned Range \\
        \hline
        \hline
        $m_{H_{2}}$~(GeV) & $\left[ 200-1000 \right] $\\
        \hline
        $m_{H_{3}}$~(GeV) & $\left [200-1000 \right ] $\\\hline$\sin{\alpha_{1}}$& $\left [(-1.0)- (1.0)\right]$\\\hline$\sin{\alpha_{2}}$& $\left[(-1.0)- (1.0) \right]$\\\hline$\sin{\alpha_{3}}$
        &$\left[(-1.0) - (1.0) \right]$\\\hline
        $m_{\chi}$~(GeV)&$\left[150-1000 \right]$ \\\hline$v_{s}$~(GeV)&$\left[150-1200\right]$\\\hline$v_{\phi}$~(GeV)&$\left[150-1000 \right]$\\\hline
        
    \end{tabular}
    \caption{\it Model parameter scan range.}
    \label{paramscanrange}
\end{table}
 
$\bullet$ {\it Stability:} The scalar potential as specified in Eq.(\ref{eq:potential}) should be bounded from below in any of the consecutive field directions. This poses constraints on the scalar quartic couplings present in Eq.(\ref{eq:potential}) and various combinations of them and can be found by ensuring the co-positivity of the mass-squared matrix for $h,s,\chi,\phi$ fields. We detail the necessary conditions that ensure the stability of the scalar potential in Appendix \ref{app:stability}. 
    
$\bullet$ {\it Perturbative Unitarity:} The couplings of the scalar potential can be constrained by perturbative unitarity of tree level $2 \to 2$ scalar scattering amplitudes. In the present framework, we have sixteen neutral and five singly charged combinations of two-particle states. We obtain the unitarity constraints and detailed derivations of the same are shown in Appendix \ref{app:PEB}.

$\bullet$ {\it Electroweak precision parameters:} The singlet scalar fields with non-vanishing mixing with the SM Higgs can provide non-trivial quantum corrections to the gauge boson self-energy diagrams. Since the extra scalars are neutral, they do not modify the $\gamma Z$ and $\gamma\gamma $ self energies ($\Pi_{\gamma Z}$ and $\Pi_{\gamma \gamma}$ respectively), only $W^\pm$ and $Z$ boson's self energies ($\Pi_{WW}$ and $\Pi_{ZZ}$) will be modified in this scenario. These effects can be easily parametrized in terms of electroweak oblique parameters, the $S$, $T$, and $U$ parameters \cite{PeskinSTU}. In the present framework, the $S$, $T$, and $U$ parameters is expressed as function of various model parameters:

 \begin{align}
   & S = \frac{1}{24\pi} \Bigl\{ \left[ \log(m_{H_1}^2) + G(m_{H_1}^2,m_Z^2) \right] (c^2_{\alpha_1} c^2_{\alpha_2} -1) + \left[ \log(m_{H_2}^2) + G(m_{H_2}^2,m_Z^2) \right] s^2_{\alpha_1}c^2_{\alpha_2} \nonumber \\&\hspace{6cm}+ \left[ \log(m_{H_3}^2) + G(m_{H_3}^2,m_Z^2) \right] s^2_{\alpha_2} \Bigr\}\\
   &  T = \frac{3}{16\pi s_W^2 m_W^2} \Bigl\{ \left[F(m_Z^2,m_{H_1}^2) - F(m_W^2,m_{H_1}^2) \right] (c^2_{\alpha_1} c^2_{\alpha_2} -1)  \nonumber \\&\hspace{1cm}+ \left[ F(m_Z^2,m_{H_2}^2) - F(m_W^2,m_{H_2}^2) \right] s^2_{\alpha_1}c^2_{\alpha_2} + \left[ F(m_Z^2,m_{H_3}^2) - F(m_W^2,m_{H_3}^2) \right] s^2_{\alpha_2} \Bigr\}\\
   & U = \frac{1}{24\pi} \Bigl\{ \left[ G(m_{H_1}^2,m_Z^2) - G(m_{H_1}^2,m_W^2) \right] (c^2_{\alpha_1} c^2_{\alpha_2} -1) \nonumber \\ & \hspace{1cm} + \left[ G(m_{H_2}^2,m_Z^2) - G(m_{H_2}^2,m_W^2) \right] s^2_{\alpha_1}c^2_{\alpha_2} + \left[ G(m_{H_3}^2,m_Z^2) - G(m_{H_3}^2,m_W^2) \right] s^2_{\alpha_2} \Bigr\}
\end{align}

where we define,
\begin{align}
 G\left(I,J\right) = -\frac{79}{3} + 9\frac{I}{J} -2\frac{I^2}{J^2} + \left(-10+18\frac{I}{J}-6\frac{I^2}{J^2}+\frac{I^3}{J^3} -9\frac{I+J}{I-J}\right) \log\left(\frac{I}{J}\right) \nonumber \\ +\left(12-4\frac{I}{J}+\frac{I^2}{J^2} \right) \frac{f(I,I^2-4IJ)}{J}\,\,, 
\end{align}

\begin{align}
 f(t,r) = 
  \begin{cases} 
   \sqrt{r} \log{ \left| \frac{t - \sqrt{r}}{t + \sqrt{r}} \right| } & \text{if } r > 0 \\
   0       & \text{if } r= 0 \\
   2 \sqrt{-r} \arctan{ \frac{\sqrt{-r}}{t}} & \text{if } r < 0\,\,,
   \end{cases}
  \end{align}

\begin{align}
 F(I,J) = 
  \begin{cases} 
   \frac{I+J}{2} - \frac{IJ}{I-J}\log(\frac{I}{J}) & \text{if } I \neq J \\
   0       & \text{if } I = J\,\, ,
  \end{cases}
\end{align}

\noindent Only new physics contributions to the $\Pi_{WW}$ and $\Pi_{ZZ}$ are included in the above expressions. The SM contributions are subtracted off, so $S=T=U=0 $ corresponds to the standard model value. A global fit of electroweak precision observable measurements provides the corresponding values of the deviations in the oblique parameters relative to their SM predictions \cite{EWPO1,EWPO2}:
\begin{eqnarray}
S_{\rm exp}=0.05\pm 0.08,\,\,T_{\rm exp}=0.09\pm 0.07,
\end{eqnarray}
for the SM Higgs mass $M_h = 125$ GeV and $U = 0$ , and correlation $\rho_{ST} = +0.92 $. For $U=0$, the electroweak oblique parameters are constrained by the following inequality \cite{Lee:2012jn} :
\begin{eqnarray}
    \frac{(S-S_{\rm exp})^2}{\sigma_S^2}+\frac{(T-T_{\rm exp})^2}{\sigma_T^2}-2\rho_{ST}\frac{(S-S_{\rm exp})(T-T_{\rm exp})}{\sigma_S\sigma_T}< R^2(1-\rho^{2}_{ST}),
\end{eqnarray}
where $\sigma_S^2$ and $\sigma_T^2 $ are one sigma experimental errors on the measured values of $S$ and $T$ respectively. 
$R^2=2.3\,,4.61\,,5.99 $ and $9.21$ for EW precision constraints at $68\%, 90\%, 95\% $ and $99\%$ confidence (CLs) limits respectively. All our benchmark points are consistent at the $95\%$ C.L.

$\bullet$ {\it Correction to $W$-boson mass:} The non-vanishing mixing between the SM and BSM Higgs sector radiatively contributes to the W boson mass. 
In particular, one loop correction to $W$ boson mass can be expressed as follows \cite{Burgess:1993mg} :
\begin{eqnarray}
    m_W = m_W^{SM} \Bigg \{ 1+ \frac{\alpha_{em}}{2 (c_W^2-s_W^2)} \left(-\frac{1}{2}S + c_W^2 T\right)\Bigg \}.
\label{mwmass}
\end{eqnarray}

We have simplified  Eq.(\ref{mwmass}) by setting $U=0$ due to its negligible contributions. $\alpha_{em} \equiv e^2/{4\pi}= g^2 s_W^2/{4\pi}$ is the fine-structure constant, $s_W (c_W) \equiv \sin \theta_W (\cos \theta_W) $ , with $\theta_W$ being the Weinberg angle. We use $m_W^{SM}=80.377\pm 0.012$\ GeV \cite{ParticleDataGroup:2022pth} and allow our model parameter space to be within $95\%$C.L. of the $m_W^{SM}$ value.  

$\bullet$ {\it Higgs observables:} In the extended Higgs sector, properties of the SM like Higgs should remain intact. In particular, we impose that the lightest mass eigenstate $(H_1)$ of the three CP-even scalars is identified with the measured SM Higgs with a mass $\approx 125$ GeV and its SM couplings \cite{PhysRevD.90.052004, CMS:2013fjq}. In any BSM physics, Higgs signal strengths are used to measure deviations in the production and decay rates of the Higgs boson from the predictions made by the SM. Deviations from the SM predictions for Higgs boson signal strengths can indicate the presence of new physics phenomena. Conversely, precise measurements of these signal strengths can be used to constrain the parameters of BSM scenarios. In this analysis, we will focus on the latter role of Higgs signal strengths to constrain BSM parameters. In our scenario, the production cross-sections for $H_1$ are suppressed by a factor of $(\cos\alpha_1 \cos \alpha_2)^2$, while the branching ratios remain unchanged. Therefore, one can define the signal strength for each production mode $i$ and decay chain $i \to H_1 \to f$ as :
\begin{eqnarray}
    \mu^f_i = \frac{[ \sigma_i(pp \to H_1) \times {\rm BR}( H_1 \to f) ]^{\rm BSM} }
    {[\sigma_i (pp \to H_{1})\times {\rm BR}(H_{1} \to f)]^{\rm SM}} = \cos^2\alpha_1 \cos^2\alpha_2, 
\end{eqnarray}

where the superscript SM indicates the SM values and the numerator is calculated in this scenario. One can impose constraints on the mixing angles, $\alpha_1$, and $\alpha_2$ through a $\chi^2 $ fit to the measured signal strengths. Similarly, one can also constrain the heavier scalar masses and the corresponding mixing angles. The fermion and gauge bosons couplings with $H_3$ are suppressed by $\sin\alpha_2$. Therefore, the production cross-section and partial decay widths of $H_3$ can be approximated as 
\begin{eqnarray}
    \sigma (pp \to H_3) &\approx & (\sin\alpha_2)^2 \sigma_{\rm SM} (pp \to H_3)\\ \nonumber 
    \Gamma( H_3 \to \psi_{\rm SM}) &\approx &(\sin\alpha_2)^2 \Gamma_{\rm SM} (H_3 \to \psi_{\rm SM}),
\end{eqnarray}

where $\sigma_{\rm SM}$ and $\Gamma_{\rm SM}$ denote the SM Higgs boson production and decay rates at the mass $m_{H_3}$ and $\psi_{\rm SM}$ represent the SM fermions and gauge bosons. In addition to these decay modes, $H_3 \to H_1 H_1 , H_1H_2,~~{\rm or}~~H_2H_2 $ are also possible if kinematically allowed. The coupling of $H_2$ to the SM particles is doubly suppressed by the factor $\sin\alpha_1 \cos\alpha_2$ and considering the mass hierarchy in this framework, $H_2$ can be produced from the decays of $H_3$, when it is kinematically accessible. To extract constraints from the SM-like Higgs boson properties we use \textsf{HiggsSignals 2.2.3} \cite{HiggsSignals2}, while to put limits on mixing angles and masses of heavier scalars we use \textsf{HiggsBounds 5.3.2} \cite{Bechtle:2020pkv}. The values of heavy scalar masses $(m_{H_2}~\&~m_{H_3})$ and three mixing angles $(\alpha_1, \alpha_2, \alpha_3)$ quoted in our benchmark points satisfy the observed limits at $95\%$ CL.

\section{One Loop Finite Temperature Effective Potential} \label{phase transition}
To study the dynamics of the electroweak phase transition in the early universe, we use the one-loop corrected finite temperature effective potential involving the SM and extra scalar fields of this scenario. We start the proceedings, by adding the Coleman-Weinberg potential $V_{\rm CW}$ and counterterms $V_{\rm CT}$ that encode one-loop corrections at zero temperature to the tree-level potential $V_0$. Finally, to incorporate the effect of temperature of the early universe, we include the finite-temperature corrections $V_{\rm T}$. The complete effective potential is given by 
\begin{eqnarray}
    V_{\rm eff} = V_{\rm 0} + V_{\rm CW} + V_{\rm CT} + V_{\rm T}
\end{eqnarray}
We first compute the Coleman-Weinberg potential $V_{\rm CW}$ using the $\overline {\rm MS}$ scheme and taking the Landau gauge to decouple any ghost contributions, and write the $V_{\rm CW}$ at zero temperature as \cite{PhysRevD.7.1888}:
  \begin{equation}
       V_{\rm cw}(H_i) = \sum_j (-1)^{F_j}\frac{n_{j}m^4_{j}(H_{i})}{64\pi^2}
\left[\log\frac{m_{j}^{2}(H_{i})}{4\pi \mu^{2}} - C_{j}  \right]
  \end{equation}
  
\noindent where, $i =1,2,3$, and $j$ runs over all particles contributing to the one-loop correction, $F_j = 0~(1)$ for bosons (fermions), $n_{j}$ are the degrees of freedom of the $j^{\rm th}$ particle and $m^2_{j}(H_i)$ is $H_i$ field dependent mass of $j$-th particle. The renormalization scale $\mu$ is fixed at $\mu = m_t (=173.5~{\rm GeV}) $, the constant $C_{j}$ is equal to $\frac{3}{2}$ for scalars and fermions and to $\frac{5}{6}$ for vector bosons. Now, taking into thermal effects, the temperature-dependent part of the effective potential at finite temperatures is given by \cite{PhysRevD.9.3320}
  \begin{eqnarray}
 V_{T}(H_{i},T)=\frac{T^4}{2\pi^2}\Bigg\{\sum_{B}n_{B}J_{B}(m_{B}^2(H_{i})/T^2) + \sum_{F}n_{F}J_{F}(m_{F}^2(H_{i})/T^2) \Bigg \}
   \label{finiteT}\end{eqnarray}
The $n_{B/F}$ are the degrees of freedom of bosons/fermions respectively and the $J_{B/F}$ are Bosonic and Fermionic functions which are represented as 
    \begin{equation}
        J_{B/F}(x^2) = \int_{0}^{\infty} y^2 \log[1\mp e^{-\sqrt{x^2+y^2}}]dy 
    \end{equation}

At high temperature limit, one can expand the Bosonic and Fermionic integrals in powers of $x \equiv m/T$ as  \cite{Quiros:1999jp}
 \begin{eqnarray}
     J_{\rm B}(x^2)\mid_{x \ll 1} &\simeq & -\frac{\pi^4}{45} + \frac{\pi^2}{12} x^2 -\frac{\pi}{6}x^3 + {\cal O}(x^4) ,\nonumber\\
     J_{\rm F}(x^2)\mid_{x \ll 1} &\simeq & \frac{7\pi^4}{360} - \frac{\pi^2}{24} x^2  + {\cal O}(x^4).
 \label{JbJf}
 \end{eqnarray} 
 In the $(m/T)$ expansion at second order, the only correction to the tree-level potential stems from the temperature-dependent nature of the mass parameters, which scale as $T^2$. Therefore, in this limit, thermal corrected one-loop effective potential reduces to a simpler form:
\begin{equation}
\label{vthigh}
    V_{T}(H_{i},T) = \frac{T^2}{24}\sum_{j} n_{j}m_{j}^2(H_{i})
\end{equation}
 
At higher temperatures, the perturbative expansion of the effective potential can lose its validity due to the appearance of infrared divergences and hence the breakdown of loop expansion \cite{Linde:1980ts,PhysRevD.9.3320}. Infrared divergences emerge from the Matsubara modes which are used to take into consideration the periodicity in the imaginary time direction in finite temperature calculations \cite{Matsubara:1955ws}. Under these circumstances, the infrared divergences can be managed by the resummation technique \cite{PhysRevD.45.4695}. This involves substituting the field-dependent masses with their respective thermal masses in the one-loop propagator. Consequently, beyond the one-loop thermal corrections to the effective potential, contributions from ring diagrams must also be included, necessitating the use of Daisy resummation. To achieve this in our analysis, we employ the Parwani method \cite{PhysRevD.45.4695} to expand the effective potential as \cite{Quiros:1992ez}: 

\begin{equation}
    V_{\rm ring}(H_{i},T)=-\sum_{j}\frac{n_{j}T}{12\pi}\Big(\big[m_{j}^2(H_{i})+\Pi_{j}(T)]^{\frac{3}{2}}-m_{j}^3(H_{i})\Big )
\end{equation}

The quantities ($m_{j}^2(H_{i})+\Pi_{j}(T))$ represent the thermal masses, where $\Pi_{i}$ are referred to as the Daisy coefficients, which are derived from the coefficients of $T^2$ terms in the finite temperature correction to the effective potential at high temperatures.

The second derivative of Eq.(\ref{vthigh}) with respect to the CP-even scalar fields yields the Daisy coefficient matrix for 
the CP-even scalar fields  $\Pi_{i}$ which is expressed as 
\begin{equation}
   \begin{bmatrix}
        \frac{\lambda_{H}}{4}+\frac{\lambda_{HS}}{12}+\frac{\lambda_{H\Phi}}{12}+\frac{3g^2}{16}+\frac{g'^2}{16}+\frac{y^2}{4} &0&0\\
    0& \frac{\lambda_{S}}{3}+\frac{\lambda_{HS}}{24}+\frac{\lambda_{S \Phi}}{12} &0\\
    0&0&\frac{\lambda_{H\Phi}}{12}+\frac{\lambda_{S \Phi }}{6}+\lambda_{\Phi}
    \end{bmatrix} T^2
\end{equation}
It is noteworthy that, while considering Coleman-Weinberg correction, the tree-level vacuum expectation values (vevs) and masses may receive modification at zero temperature. To prevent such changes, it is necessary to incorporate counterterms into the effective potential. The counterterm potential is defined as:
\begin{align}
 \delta V_{\rm ct}(H_{i}) =\,& {\delta\mu_{H}^2}\,( H^\dagger H)+{\delta\lambda_{H}}\, 
( H^\dagger H)^2 +{\delta\mu_{\Phi}^2}\,\Phi^2 + \delta\lambda_{\Phi} \,  \Phi^4 + {\delta\mu_{S}^2}\,( S^\dagger S) +\delta\lambda_{S} \, ( S^\dagger S)^2+ \nonumber \\& \delta\lambda_{H S}\,
(H^\dagger H)(S^\dagger S)+\delta\lambda_{S \Phi}(S^\dagger S) \Phi^2+\delta\lambda_{H \Phi}(H^\dagger H)\Phi^2 +\frac{\delta\mu_3}{2}(S^3+S^{\dagger3}) 
\label{ctpot}
\end{align}

To determine the expressions for the counterterms corresponding to each parameter, we use the following conditions:
\begin{eqnarray}
    \partial_{h_{a}}(\delta V_{\rm ct}+\Delta V)=0 \nonumber\\
     \partial_{h_{a}}\partial_{h_{b}}(\delta V_{\rm ct}+\Delta V)=0,
     \label{ctcon}
\end{eqnarray}
where the partial derivatives are taken with respect to $h$ , $s$ and $\phi$ fields expressed as $h_{a/b}$. The derivatives are evaluated at the vacuum expectation values of the respective fields, $\langle H\rangle =v_{h}$, $\langle S\rangle=v_{s}$, $\langle\Phi\rangle =v_{\phi}$, and $\Delta V$ is the effective potential at zero temperature excluding the tree-level part of the potential. The expression of each coefficient of counter terms is shown in Appendix \ref{AppctN}. The final expression of the total effective potential is given by 
\begin{equation}
    V(H_{i},T) = V_{0}(H_{i})+ V_{\rm cw}(H_{i})+V_{T}(H_{i},T)+V_{\rm ring}(H_{i},T)+\delta V_{\rm ct}(H_{i})
    \label{Vtot}
\end{equation}

The critical temperature $T_{c}$ in a phase transition is determined by setting the potential values at two vacuum expectation values (vevs), namely the High vev and the Low vev, equal to each other. The existence of a barrier between the symmetric phase (high temperature) and the broken phase (low temperature) is the requirement for the strong first order phase transition (SFOPT). This can be established using the following mathematical expression \cite{Patel:2011th}.
   \begin{equation}
    V(H_{i}^{\rm High},T_{c})= V(H_{i}^{\rm Low},T_{c})
\end{equation}

We define the order parameter $\zeta_{c,i}$ along the $i^{th}$ field direction as:
\begin{equation}
    \zeta_{c,i} = \frac{\Delta H_{i}}{T_{c}}
\end{equation} with $\Delta H_{i}$ is the difference of high and low vevs of the SM/BSM scalar field. The criterion for an SFOPT is given by $\zeta_{c, i} \geq 1$. 
It is important to recognize that the overall effective potential in Eq.(\ref{Vtot}) explicitly depends on the gauge choice. Consequently, the extrema of the effective potential and the order parameter $\zeta_c $ are gauge dependent~\cite{PhysRevD.9.3320, Patel:2011th, PhysRevD.51.4525, Garny:2012cg}. In our analysis, all finite temperature calculations are performed in the Landau gauge ($\xi=0$)\footnote{It is important to note that one can construct gauge independent variables using Nielson's identities from the effective potential  \cite{Nielsen:1975fs, PhysRevD.13.3469}. Detailed analysis is beyond the scope of this paper.}.  Due to the presence of the $\mu_{3}$ term in the tree-level potential, barrier formation takes place at the tree level which facilitates FOPT. Therefore, the gauge dependence that appears through one-loop-induced corrections does not significantly affect our results. Such studies have been conducted in the literature \cite{Borah:2023zsb,Ghosh:2022fzp}.

A first-order phase transition involves two key temperatures: the critical temperature ($T_c$) and the bubble nucleation temperature ($T_n$). Generally, the transition proceeds through bubble nucleation at $T_n$, which is typically below $T_c$. During this process, at a finite temperature $T$, the probability per unit volume of tunneling from the false vacuum to the true vacuum is given by, \cite{PhysRevD.75.043507}
\begin{equation}
    \Gamma(T) = T^4 \left(\frac{S_{3}}{2\pi T}\right)^{\frac{3}{2}} e^{-\frac{S_{3}}{T}} 
\end{equation}

where $S_3$ represents the 3-dimensional Euclidean action and is represented as \cite{1983544}
\begin{equation}
    S_{3}=4\pi \int  r^2 dr\left[\frac{1}{2}\left(\frac{d H_{i}}{dr}\right)^2 + V_{\rm eff}(H_{i},T)\right]
\end{equation}

where the scalar field $H_i$ follows the differential equation given by \cite{1983544,PhysRevD.45.3415,PhysRevLett.46.388}
\begin{equation}
    \frac{d^2H_{i}}{dr^2} + \frac{2}{r}\frac{dH_{i}}{dr}=\frac{dV_{\rm eff}(H_{i},T)}{dH_{i}}
\end{equation}

with the boundary conditions $H_{i}=0$ as $r\rightarrow\infty$ and $\frac{dH_{i}}{dr}=0$ at $r=0$.

The nucleation temperature is then defined as the temperature at which the following condition is satisfied \cite{PhysRevD.45.3415,1983544}.
\begin{equation}
    \frac{S_{3}(T_{n})}{T_{n}}=140 \label{nuctemp}
\end{equation}
\setlength{\tabcolsep}{12pt}
\begin{table}[h!]
    \centering
    \begin{tabular}{|c|c|c|c|c|c|c|c|}
      \hline 
      Parameters&  BP 1 & BP 2 & BP 3 &  BP 4 & BP 5 & BP 6 &BP 7    \\\hline \hline
    $m_{H_{2}}$ (GeV)& 377.15 & 296.93 & 260.50& 275.64 & 234.94 & 290.09 &377.22\\\hline
    $m_{H_{3}}$ (GeV)  & 536.38 & 496.55& 600.82& 331.63 & 367.80& 631.88&764.36\\\hline
    
    $m_{\chi}$ (GeV) &985.94&653.75&765.83&714.46&787.2&628.44&840.19\\\hline 
    
    $\sin\alpha_{1}$ & 0.0016&-0.067&0.112& -0.0252&-0.0086&-0.022&0.12\\\hline

    $\sin\alpha_{2}$ & 0.161&0.22&0.172& 0.189&0.196&0.229&0.06\\\hline

    $\sin\alpha_{3}$  & 0.192& 0.094&0.025 & 0.877& 0.965&0.175&-0.0074\\\hline

    $v_{s}$ (GeV)  & 391.28 & 870.91 & 915.97 & 746.70 & 851.43 & 962.96  &1013.05\\\hline
    
    $v_{\phi}$ (GeV)  & 152.16 & 121.6 &184.43& 156.33 & 227.8 &238.13&624.37\\\hline
    
    \end{tabular}
    \caption{\it Benchmark Points for the study of SFOPT.}

    \label{benchmarkpoints}
\end{table}
As we shall see later in Sec.\ref{Gravitywave}, the main ingredients to calculate the stochastic  gravitational wave spectrum from cosmological phase transitions are the quantities $\{\alpha, \frac{\beta}{H_{n}}, T_{n}\}$, where $\alpha$ is related to the latent heat evolved during the phase transition, $\beta$ is the inverse time duration of phase transition, $T_{n}$ is the nucleation temperature defined above in Eq.(\ref{nuctemp}) and $H_{n}$ is the Hubble parameter evaluated at the nucleation temperature. We generate the results of the phase structures of the scalar fields using the publicly available \textsf{CosmoTransitions} \cite{Wainwright:2011kj} package and finally evaluate the output quantities $\{\alpha, \frac{\beta}{H_{n}}, T_{n}\}$ corresponding to our benchmark points. 
\begin{table}[h]
\begin{tabular}{| c | c | c | c | c | c | c | c |} 
\hline
PT Parameters &  BP1 & BP2 & BP3 \\ \hline \hline
$T_{c}$ (GeV)  &  124.78 &  I: 674&   I: 236   \\
&&  II: 133&  II: 74\\\hline
$T_{n}$ (GeV) &     124.65 &  I: 555   &      I: 232 \\
&&  II: 127&   II: No Nucleation\\\hline
$\alpha$   & 0.071    &I : 0.0008 &  I: 0.0010  \\
&&   II: 0&\\\hline
$\frac{\beta}{H_n}$    &  $4.15\times10^{7}$ &  I: 1356.85 &     I: 22516.8  \\
&&   II: 7955.0&\\\hline
$(\langle H_1 \rangle,\langle H_2 \rangle, \langle H_3\rangle)|_{T_{c}}$ &(0,0,0)&  I: ${(0,0,0)} $ &  I:  ${(0,0,0)} $\\
(High $T_{c}$ vevs)&&  II: (0,874,0)&  II: (0,0,137)\\\hline 
$(\langle H_1 \rangle,\langle H_2\rangle, \langle H_3\rangle)|_{T_{c}}$  & (150,386,134)   &  I: (0,717,0)&    I: (0,0,112)\\
(Low $T_{c}$ vevs)&&  II: (142,872,93)&  II: (235,915,182)\\\hline
$(\zeta_{c,1}, \zeta_{c,2}, \zeta_{c,3})$&(1.2, 3.09, 1.07)&  I: (0, 1.06, 0)&  I: (0, 0, 0.47)\\
&&  II: (1.06, 0.01, 0.73)&  II: (3.17, 12.3, 0.60)\\\hline
$(\langle H_1\rangle,\langle H_2\rangle, \langle H_3\rangle)|_{T_{n}}$ & (0,0,0) &  I: (0,0,0)&    I: (0,0,0)  \\(High $T_{n}$ vevs)&&  II: (0,874,0)&\\\hline
$(\langle H_1\rangle,\langle H_2\rangle, \langle H_3\rangle)|_{T_{n}}$&(173,387,137)&  I: (0,794,0)&    I: (0,0,114) \\(Low $T_{n}$ vevs)&&  II: (162,873,98)&\\\hline

\end{tabular}
\caption {\it Phase transition output parameters for SM and BSM Higgs directions corresponding to benchmark points BP1, BP2, BP3. For each scalar field, the high and low vacuum expectation values (vevs) $(\langle H_1\rangle, \langle H_2\rangle, \langle H_3\rangle)$ are presented at both the critical temperature $T_c$ and nucleation temperature $T_n$ for Step I and Step II of the phase transition. The phase transition strength along the $H_i$ direction is denoted by $\zeta_{c,i}$, $\alpha$ represents the latent heat released, $\beta $ is the inverse duration of the phase transition, and $T_n$ signifies the nucleation temperature. Temperatures and vacuum expectation values are in GeV units.}  
    \label{PTSM}
\end{table}

We now analyze the SFOPT corresponding to the 7 benchmark points as shown in Table \ref{benchmarkpoints}.  We broadly classify the phase transition along both the SM+BSM Higgs field directions and only along the BSM Higgs field directions. In the next two subsections, we will delve into these two.
\begin{figure}[ht]
  \subfigure[]{
  \includegraphics[scale=0.3]{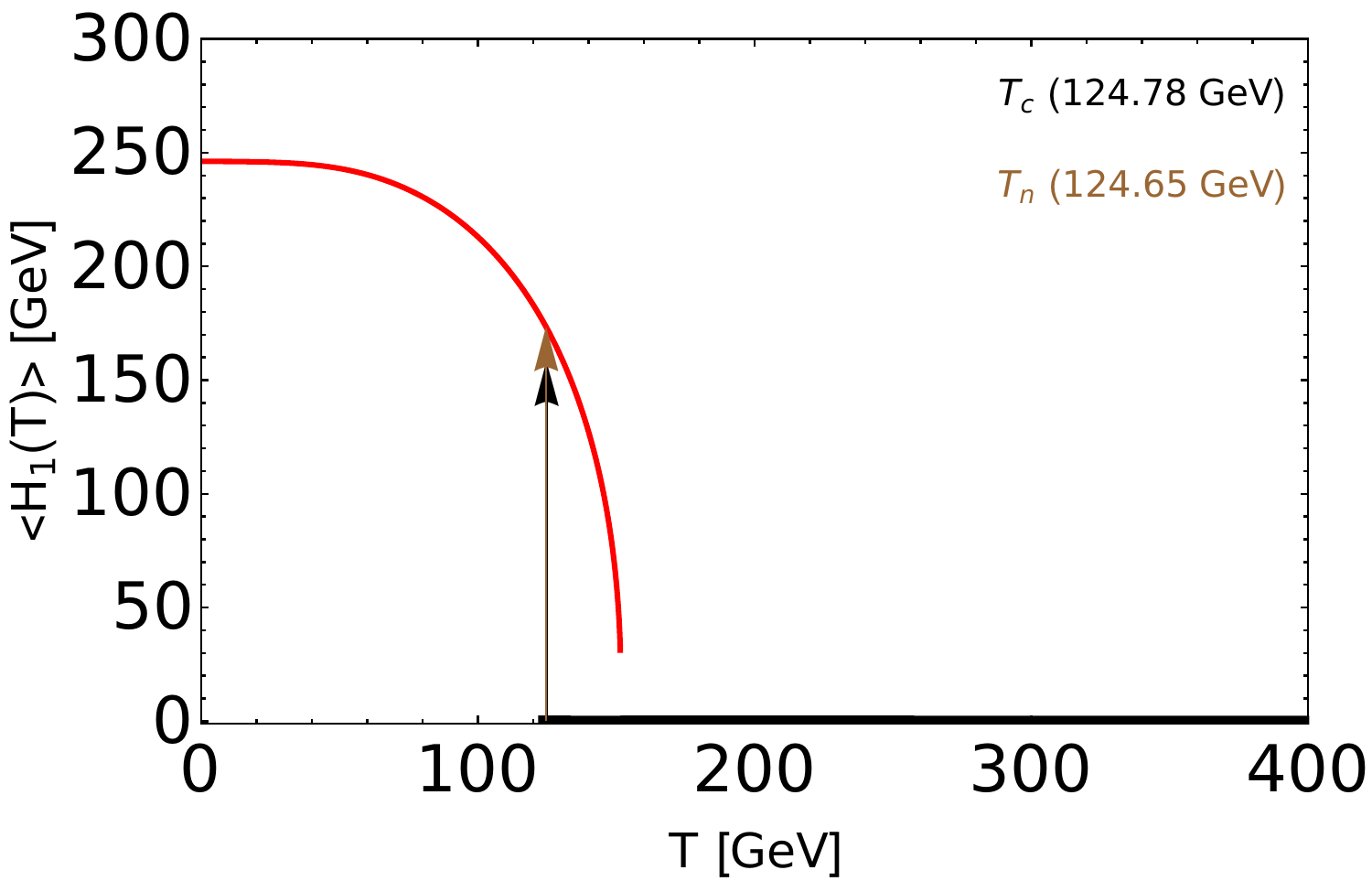}}
  \subfigure[]{
  \includegraphics[scale=0.3]{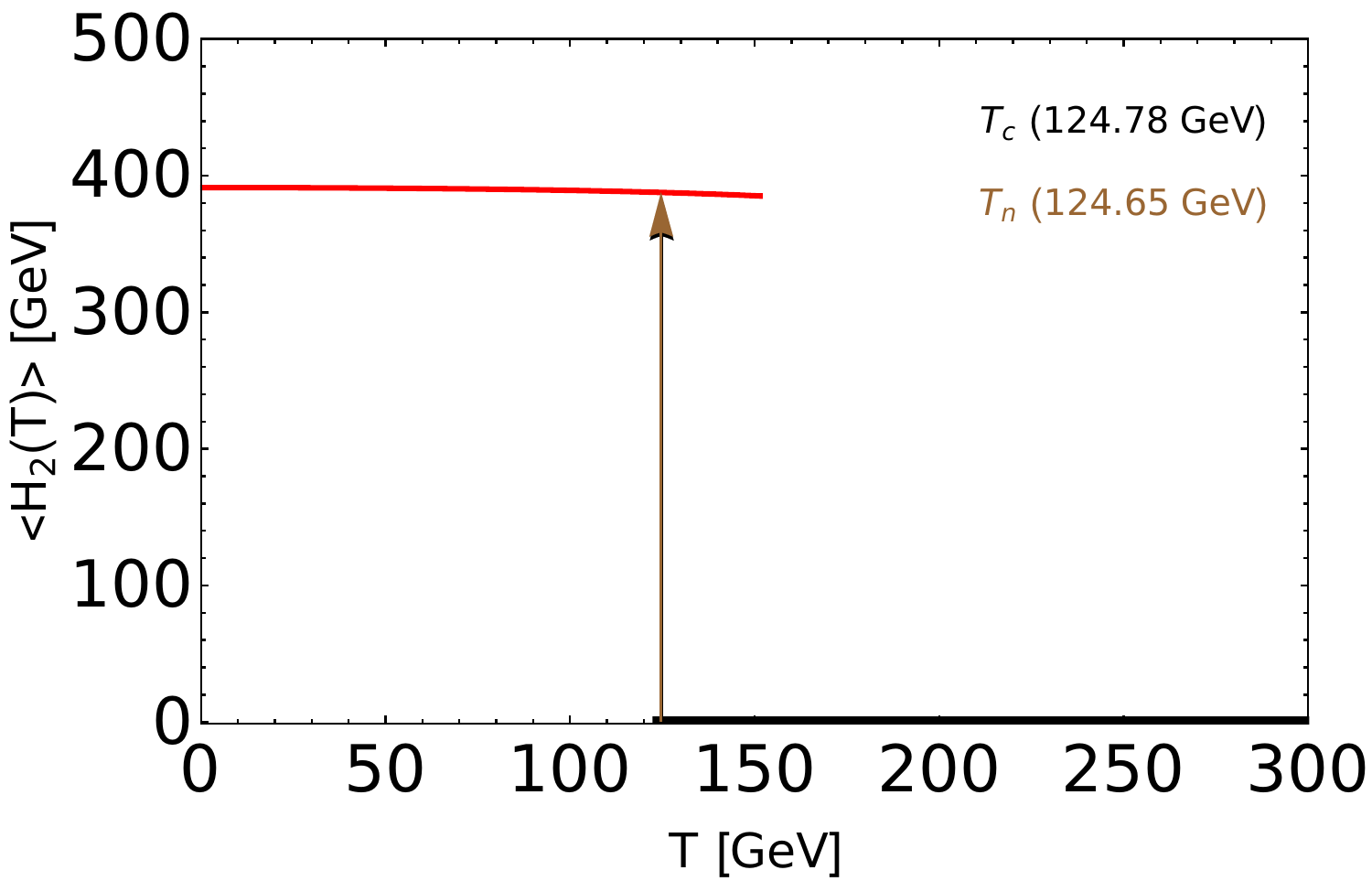}}
  \subfigure[]{
  \includegraphics[scale=0.3]{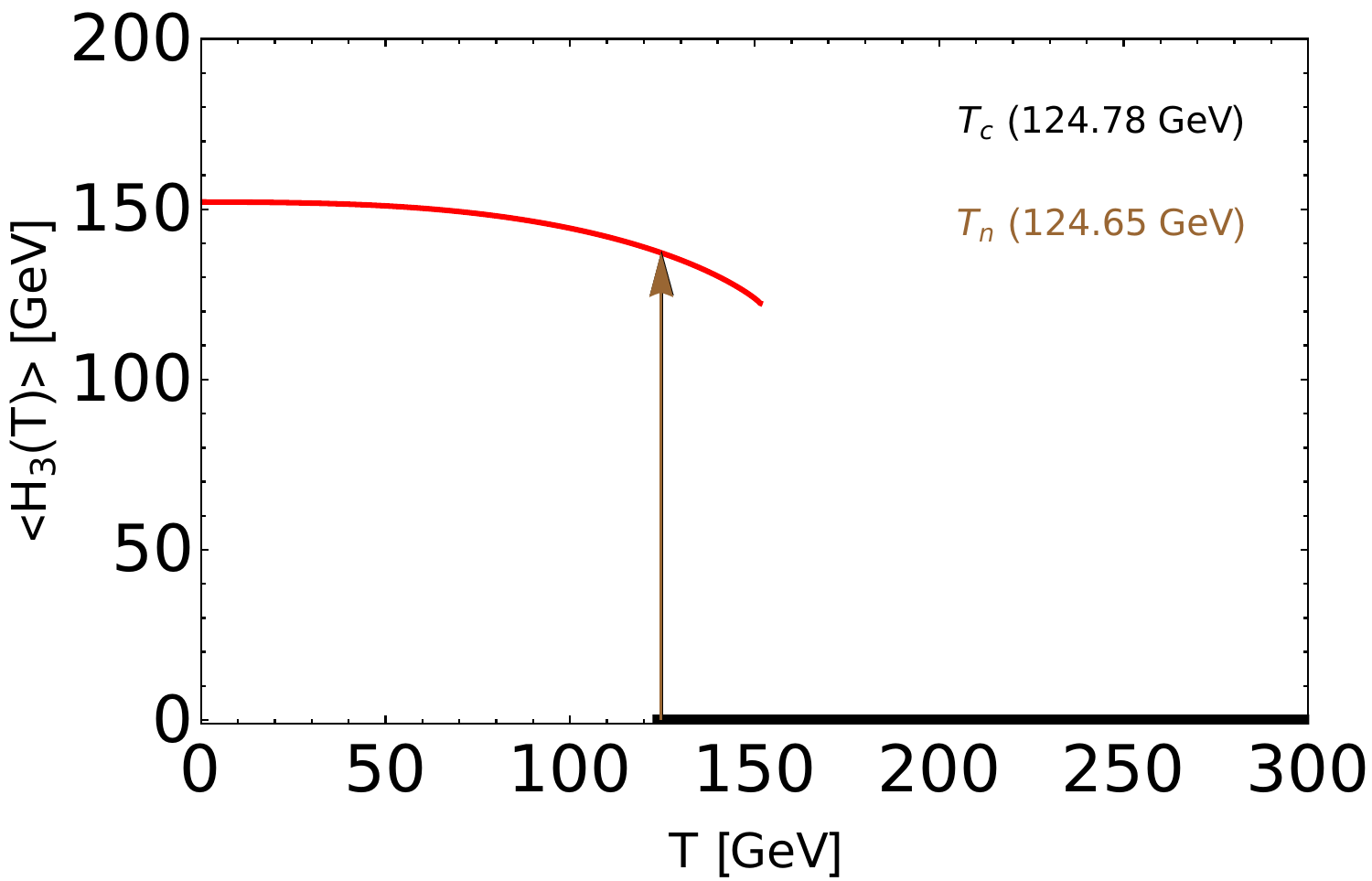}}
    \caption {\it The phase structure of the scalar fields $(H_1, H_2, H_3)$ is shown for benchmark point BP1 as a function of temperature. Figures $(a)$, $(b)$, and $(c)$ represent the evolution of the vacuum expectation values (vevs) of these fields with temperature. Both critical and nucleation temperature analyses are included. Different colours represent distinct phases, with colour changes accompanied by arrows indicating first order phase transitions and color changes without arrows indicating second order phase transitions.}
    \label{PhTFig1}
\end{figure}

\subsection{Strong First Order phase transition along SM+BSM Higgs field direction} \label{BP123}
Based on Table \ref{PTSM}, one can discern that benchmark points BP1, BP2, and BP3 exhibit strong first order phase transition along both the SM Higgs ($H_{1}$) and BSM Higgs ($H_{2}$ or both $H_{2}$ and $H_{3}$) directions. We show the phase structure of the scalar fields as a function of temperature in Figure \ref{PhTFig1}. For benchmark point BP1,  Figure \ref{PhTFig1} reveals a single step first order phase transition at the critical temperature $T_{c}=124.78$ GeV. This transition involves a shift from $\{0,0,0\}\rightarrow\{H_{1}, H_{2}, H_{3}\}$, where the quantities in the parenthesis represent the high and low vacuum expectation values (vevs) of the respective scalar fields at temperature $T_{c}$. Notably, this phase transition is strong for all field directions corresponding to this benchmark point, characterized by order parameters  $\zeta_{c,1}=1.2$, $\zeta_{c,2}=3.09$ and $\zeta_{c,3}=1.07$.

In the case of BP2 as shown in Table \ref{PTSM}, we see a two-step first order phase transition. The first transition happens at $T_{c}=645 $ GeV along the $H_{2}$ field direction, followed by a second transition at $T_{c}=133$ GeV along the three field directions. The overall transition sequence is $\{0,0,0\}\rightarrow\{0,H^{'}_{2},0\}\rightarrow\{H_{1},H_{2},H_{3}\}.$ Here, $\{0, H^{'}_{2},0\}$ represents the low temperature vevs at the first step, and  $\{H_{1}, H_{2}, H_{3}\}$ represents the low temperature vevs at the second step. In the first step, the transition along the $H_{2}$ direction is strong as $\zeta_{c,2}=1.06$. In the second step, the transition is strong only along the $H_{1}$ field direction, while it is weak along the $H_{2}$ and $H_{3}$ field directions, with  $\zeta_{c,1}=1.06$, $\zeta_{c,2}=0.01$ and $\zeta_{c,3}=0.73$. 

The phase transition pattern seen for benchmark point BP3 is similar to that of BP2, featuring a two-step first order phase transition) with the following sequence $\{0,0,0\}\rightarrow\{0,0, H^{'}_{3}\}\rightarrow\{H_{1}, H_{2}, H_{3}\}.$ Here, $\{0,0, H^{'}_{3}\}$ and $\{H_{1}, H_{2}, H_{3}\}$ represent the low temperature vacuum expectation values (vevs) at the respective stages of the phase transition. In both steps of the phase transition, the transition along the $H_{3}$ field direction is weak with order parameter values $\zeta_{c,3}=0.47$ and $0.6$ respectively. In contrast, the transitions along the $H_{1}$ and $H_{2}$ field directions
are strong in the second step, with order parameter values $\zeta_{c,1}=3.17$ and $\zeta_{c,2}=12.3$ respectively as shown in Table \ref{PTSM}.

\begin{figure}[hbt!]
  \subfigure[]{
  \includegraphics[scale=0.39]{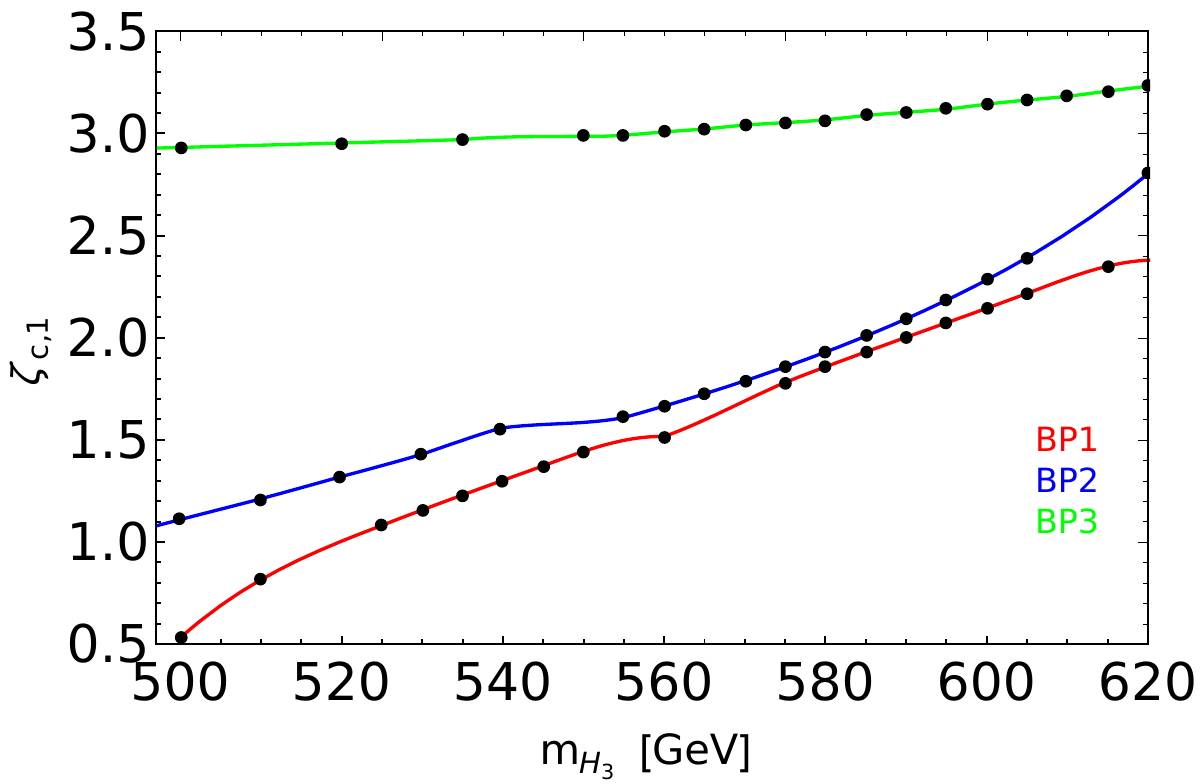}}
  \subfigure[]{
  \includegraphics[scale=0.39]{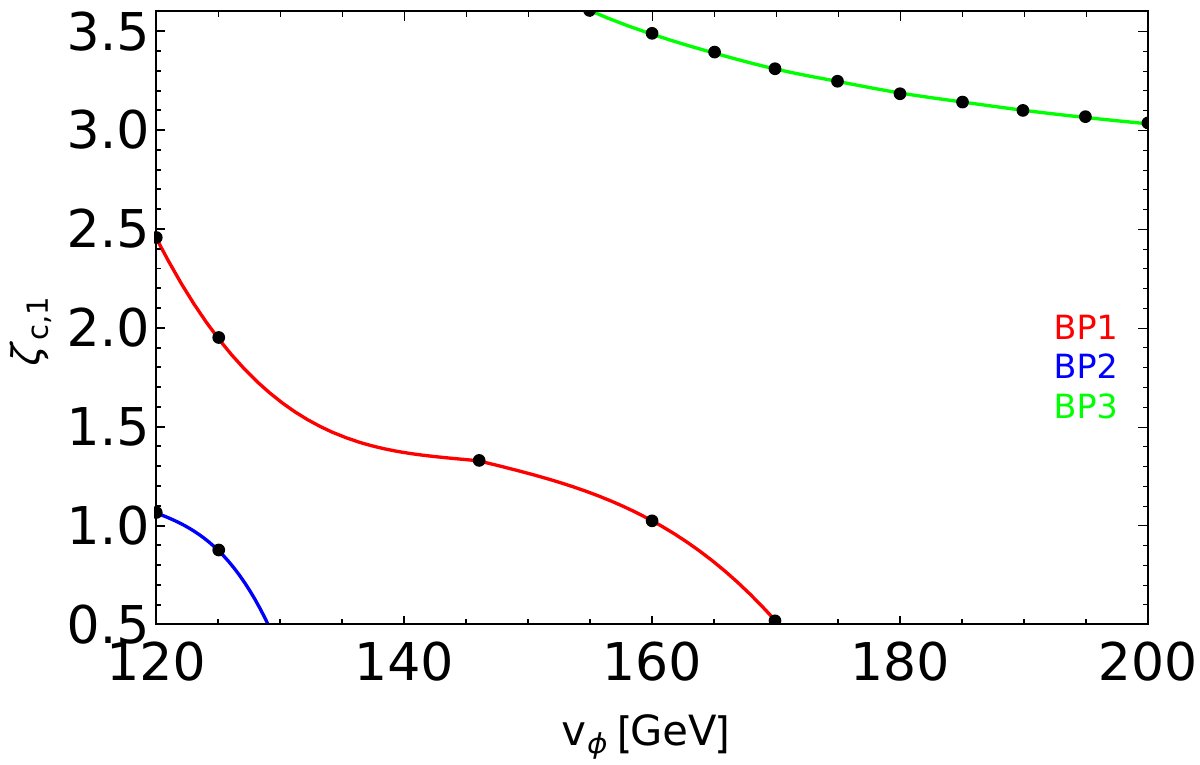}}
  \subfigure[]{
  \includegraphics[scale=0.39]{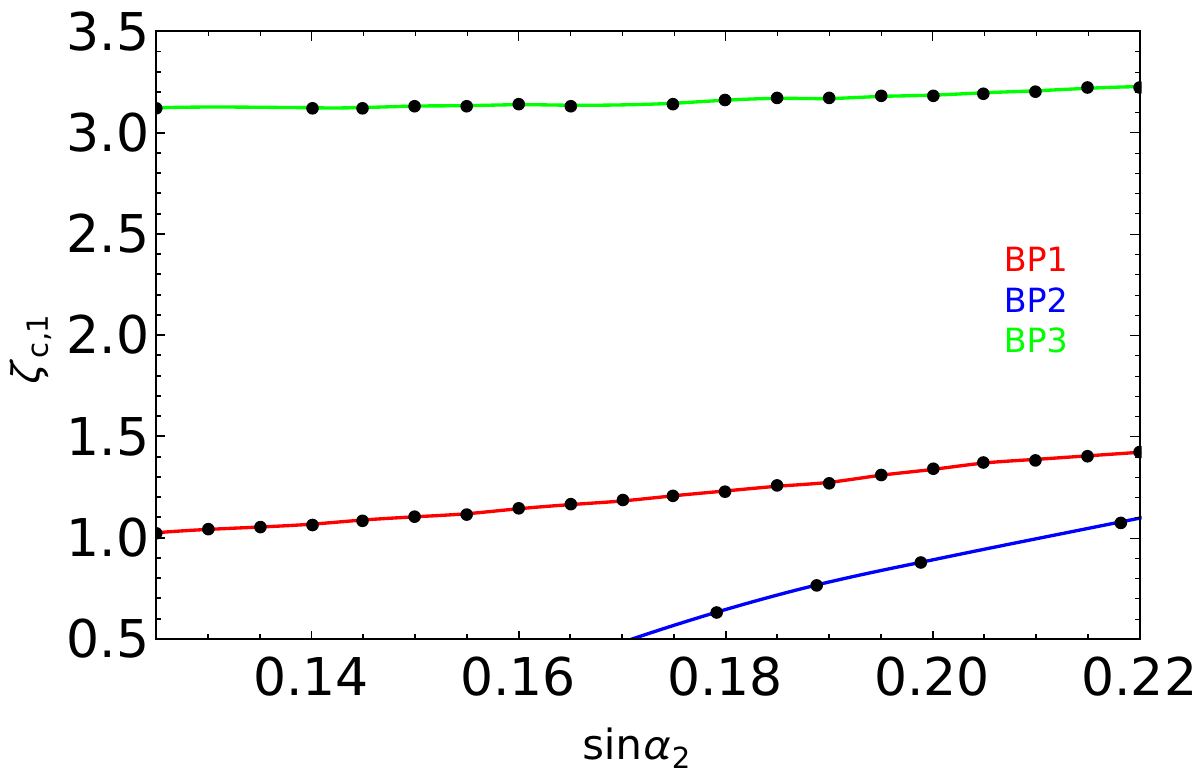}}
    \caption{\it Variation of the phase transition strength along SM Higgs field direction on three model parameters. Dots represent the exact value of order parameter obtained at the respective input variable. } 
    \label{paramrole}
\end{figure}

Having elucidated the general phase transition patterns for a representative set of benchmark points within this scenario, we now delve into the model's parametric dependence on the phase transition strength. This analysis will focus on three important model parameters: $m_{H_3}, \sin\alpha_2$, and $v_\phi$ which play a crucial role in determining the strength of the phase transition. As a representative scenario, we choose 3 benchmark points (BP1, BP2, BP3), characterized by a strong phase transition along the SM Higgs direction.

In the upper panel of Figure \ref{paramrole}, we illustrate the dependence of phase transition strength along SM Higgs field direction $\zeta_{c,1}$ on $m_{H_{3}}$. We vary $m_{H_{3}}$ for this particular analysis while keeping other model parameters unchanged for BP1, BP2~\&~BP3. The results demonstrate a positive correlation between $\zeta_{c,1}$ and $m_{H_3}$, indicating an increase in the strength of the phase transition with an increase in $m_{H_{3}}$. This trend is observed for all other benchmark points that exhibit SFOPT along the SM Higgs field direction. Similarly, an increase in $\sin\alpha_{2}$ results in increased phase transition strength, whereas the reverse effect is observed with the increase of $v_{\phi}$, as shown from the top-left, top-right, and bottom panels respectively. The influence of other model parameters on phase transition strength appears negligible. This behaviour can be understood from Eq.(\ref{qcoupling}), where the criteria of SFOPT is enhanced by increasing the scalar quartic couplings which have $m_{H_{\rm 3}}$, $\sin\alpha_{2}$ in the numerator and $v_{\phi}$ in the denominator.

\subsection{Strong First Order phase transition along BSM Higgs field direction}
\label{BP4567}
\begin{figure}[hbt!]
  \subfigure[]{
  \includegraphics[scale=0.3]{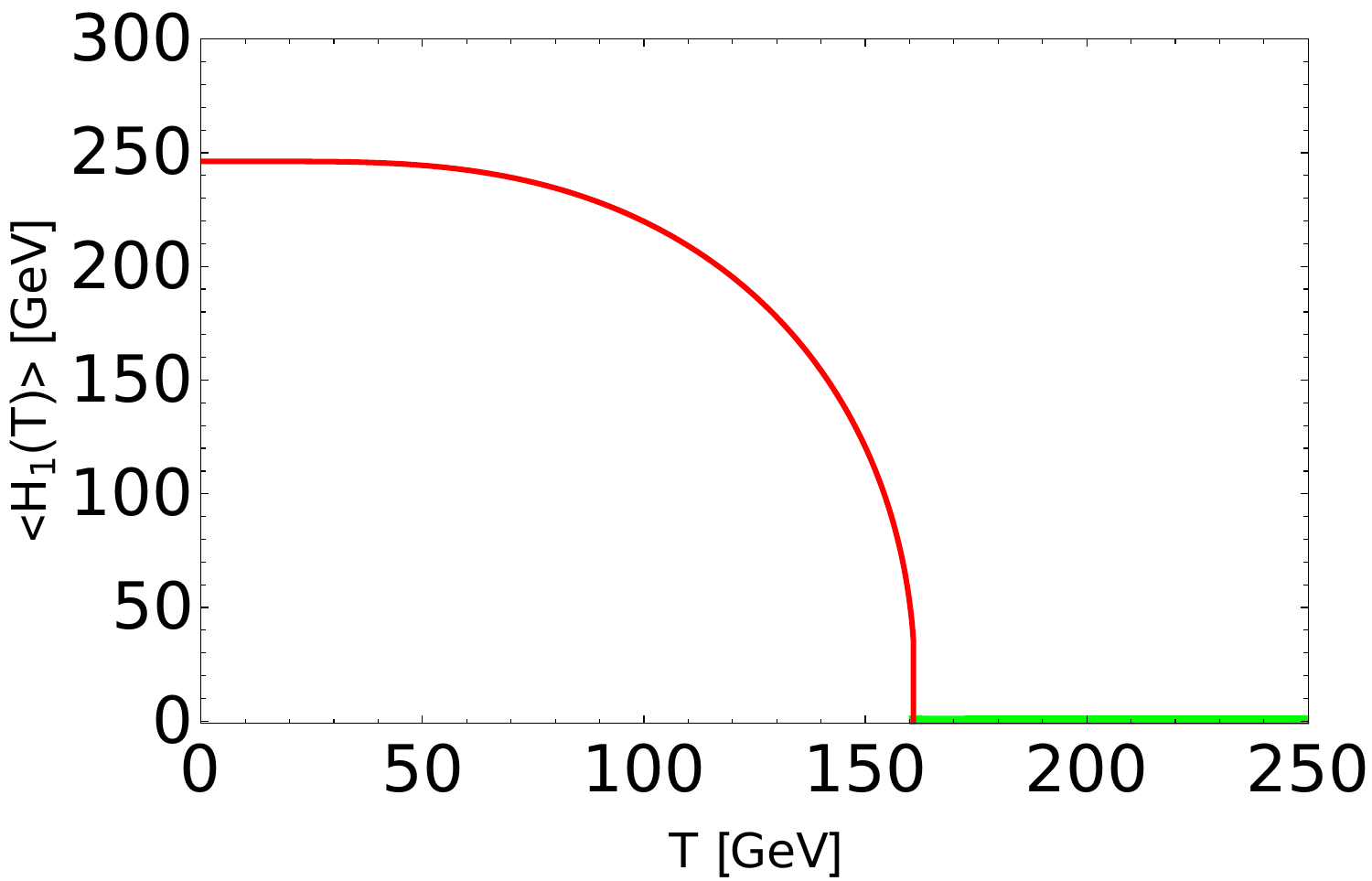}}
  \subfigure[]{
  \includegraphics[scale=0.32]{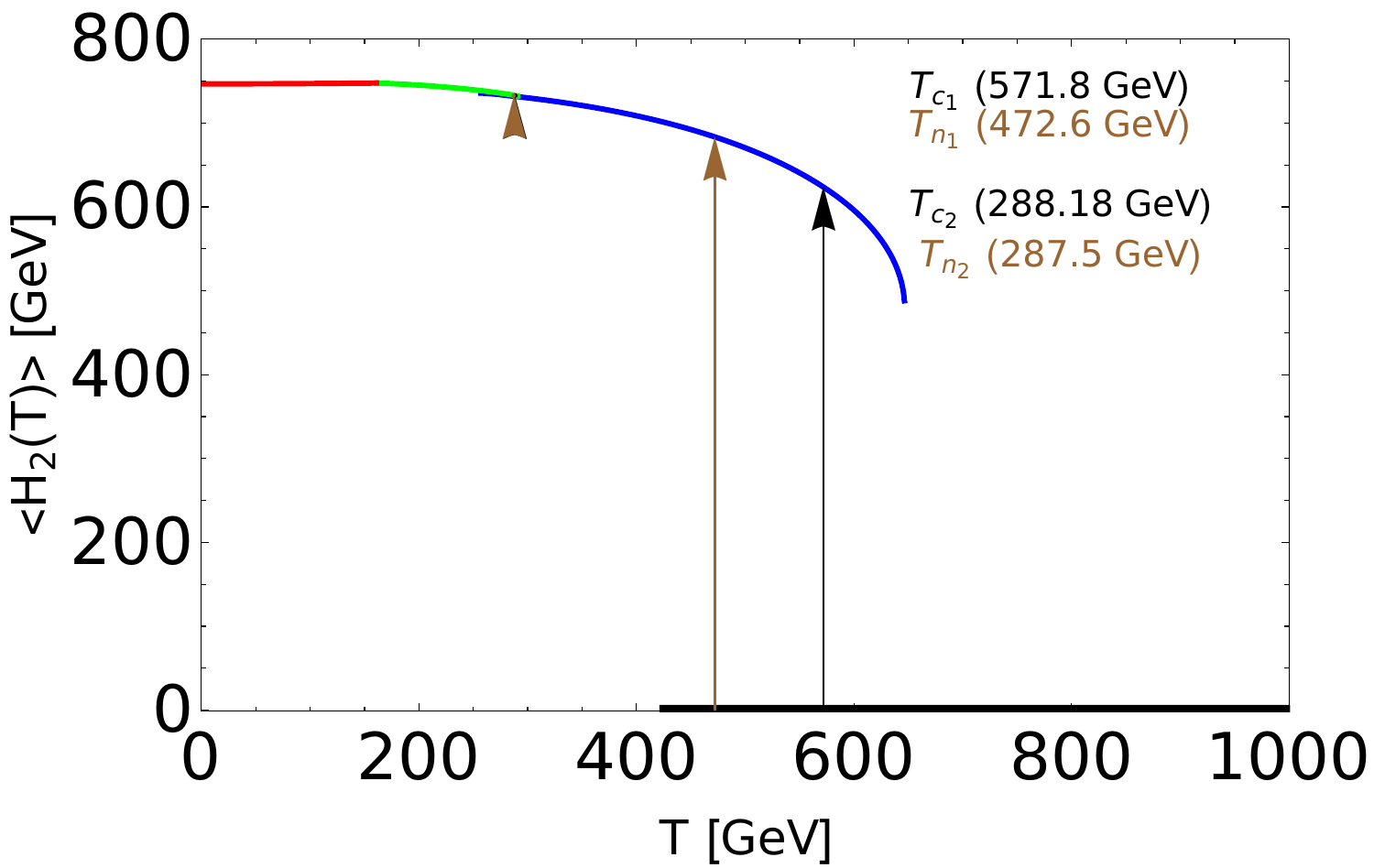}}
  \subfigure[]{
  \includegraphics[scale=0.3]{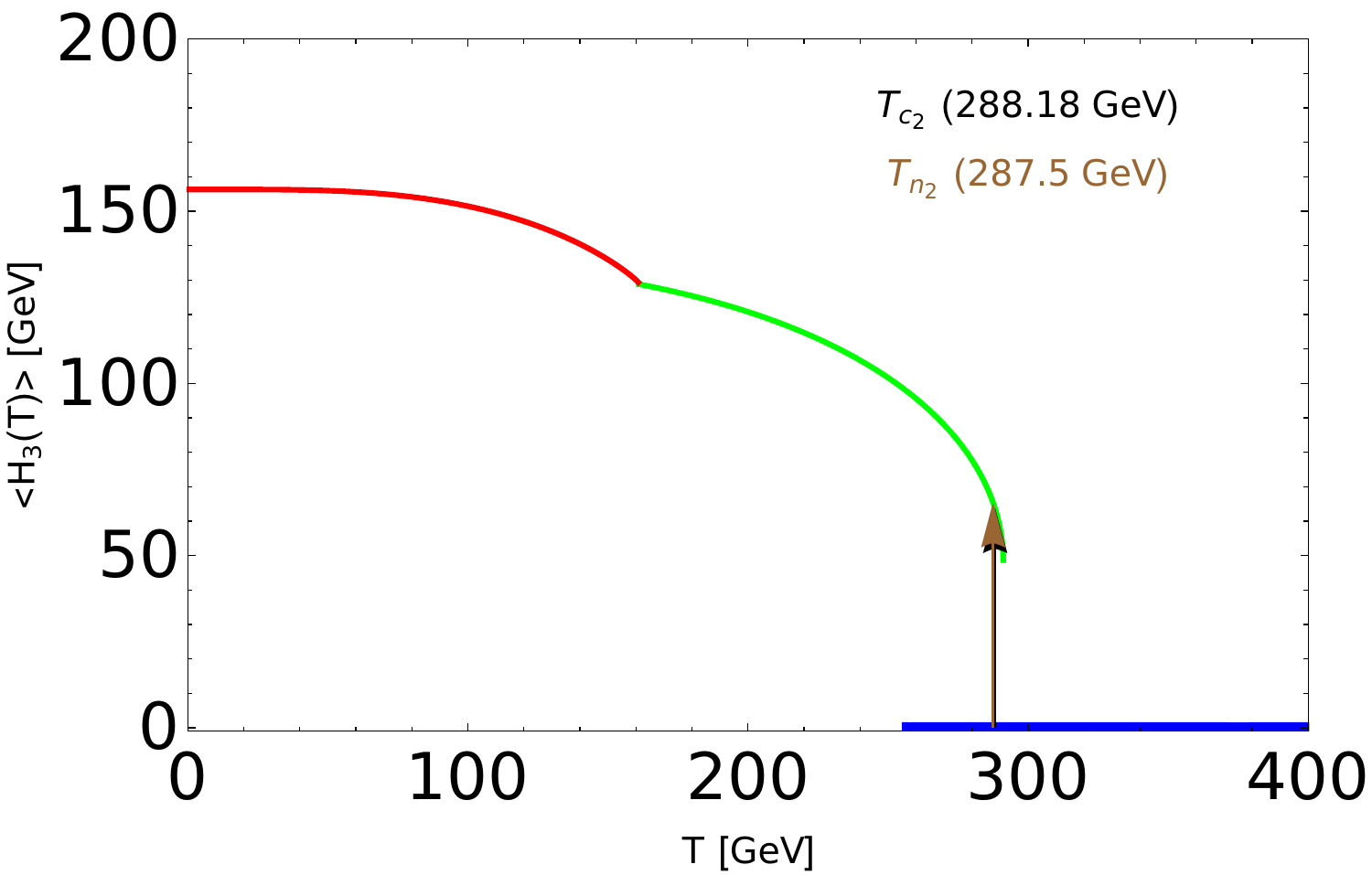}}
    \caption{\it Phase Structure of the fields as a function of temperature corresponding to BP4. (a), (b), (c) corresponds to the evolution of the vevs of $H_{1}, H_{2}, H_{3}$ scalar fields with respect to temperature. We show the results for both critical and nucleation temperature analysis. Different colours indicate different phases.  Colour change with (without) arrow indicates a first (second) order phase transition.  }
    \label{PhTFig4}
\end{figure}

Based on input model parameters for benchmark points BP4-BP7 as shown in Table \ref{benchmarkpoints}, we exhibit various output parameters of the strong first-order phase transition along the BSM Higgs $(H_2)$ direction corresponding to those benchmark points in Table \ref{PTBSM}. From this Table, we see an interesting pattern of phase transition flow. For example, we see that BP4, BP5, and BP6, represent a three-step phase transition characterized by the sequence  $\{0,0,0\}\rightarrow\{0,{H_{2}}^{''},0\}\rightarrow\{0,{H_{2}}^{'},{H_{3}}^{'}\}\rightarrow\{H_{1}, H_{2}, H_{3}\}$ where $\{0,{H_{2}}^{''},0\}$, $\{0,{H_{2}}^{'},{H_{3}}^{'}\}$ and $\{H_{1}, H_{2}, H_{3}\}$ represent the low temperature vacuum expectation values at each step of phase transition respectively. For BP4-BP6, the initial step involves a strong first order phase transition along the $H_{2}$ direction with $\zeta_{c,2}$ values 1.09, 1.07 and 1.05 at $T_{c_{1}}$ = 571.8 GeV, 654 GeV and 717 GeV respectively. In the second step, a weak first order phase transition occurs along both the $H_{2}$ and $H_{3}$ field directions with $\zeta_{c,2}$ values 0.003, 0.002, 0.03 and $\zeta_{c,3}$ values  0.22, 0.18, 0.35  at $T_{c_{2}}$ = 288.18 GeV, 409.55 GeV and 399.2 GeV for BP4, BP5 and BP6 respectively.  The final (third) step is a second-order phase transition in all the field directions.

\setlength{\tabcolsep}{8pt}
\begin{table}[h]
\begin{tabular}{| c | c | c | c | c | c | c | c |} 
\hline
PT Parameters &  BP4 & BP5 & BP6& BP7 \\ \hline \hline
$T_{c}$ (GeV) &    I: 571.8&   I: 654 &   I: 717&726\\&  II:
 288.18 &  II: 409.55&  II: 399.2&\\\hline
$T_{n}$ (GeV)&    I:  472.6&       I: 584&  I: 625&541\\&  II: 287.5 &  II: 408&  II: 395&\\\hline
$\alpha$   & I: 0.0012    &I: 0.0011& I: 0.0007 &0.0019  \\&  II: 0&  II: 0&  II: 0&\\\hline
$\frac{\beta}{H}$    &   I: 1388.73  &   I: 1537.46    &   I: 1120.58 &740.53\\&  II: 166191&  II: 150696&  II: 29075.8&\\\hline
$(\langle H_1\rangle,\langle H_2\rangle, \langle H_3\rangle)|_{T_{c}}$ &  I: (0,0,0)&  I: (0,0,0)&  I: (0,0,0)&(0,0,568)\\(High $T_{c}$ vevs)&  II: (0,732,0)&  II: (0,813,0
)&  II: (0,903,0)&\\\hline 
$(\langle H_1\rangle,\langle H_2\rangle, \langle H_3\rangle)|_{T_{c}}$ &   I:(0,624,0)&  I: (0,700,0) 
&Step I:(0,756,0)&(0,848,569)\\(Low $T_{c}$ vevs)&  II: (0,733,64) &  II: (0,814,74)&  II:  (0,916,140)&\\\hline
$(\zeta_{c,1}, \zeta_{c,2}, \zeta_{c,3})$&  I: (0, 1.09, 0)&  I: (0, 1.07, 0)&  I: (0, 1.05, 0)&(0, 1.16, 0.001)\\
&  II: (0, 0.003, 0.22)&  II: (0, 0.002, 0.18)&  II: (0, 0.03, 0.35)&\\\hline
$(\langle H_1\rangle,\langle H_2\rangle, \langle H_3\rangle)|_{T_{n}}$&   I: (0,0,0) &  I: (0,0,0) &   I:(0,0,0) &(0,0,597)\\(High $T_{n}$ vevs)&  II: (0,731,0)&  II: (0,813,0)&  II: (0,903,0)&\\\hline
$(\langle H_1\rangle,\langle H_2\rangle, \langle H_3\rangle)|_{T_{n}}$&  I: (0,683,0) &  I:(0,812,0) &  I: (0,844,0), &(0,944,598)\\(Low $T_{n}$ vevs)&  II: (0,733,65)&  II: (0,814,77)&  II:(0,918,144)&\\\hline

\end{tabular}
    \caption{\it Phase transition output parameters for SM and BSM Higgs directions corresponding to BP4-7. The values of high and low vevs $(\langle H_1\rangle,\langle H_2\rangle, \langle H_3\rangle)$ of each of the scalar fields is given at all steps for all the benchmark points for both $T_{c}$ and $T_{n}$. ${\rm I}$ represents Step ${\rm I}$ while ${\rm II}$ represents Step ${\rm II}$ in phase transition. Temperature and vevs are in GeV unit. Here the quantity $\zeta_{c,i}$ denotes the phase transition strength along the $H_{i}$ direction, $\alpha$ is related to the latent heat evolved during the phase transition, $\beta$ is the inverse time duration of phase transition, $T_{n}$ is the nucleation temperature.  }
    \label{PTBSM}
\end{table}

As alluded in Table \ref{PTBSM}, Higgs field corresponding to benchmark point BP7 undergoes a two step phase transition. The initial step is characterised as a first order while the subsequent step is a second order transition. The transition pattern is as follows : $\{0,0,{H_{3}}^{''}\}\rightarrow\{0,{H_{2}}^{'},{H_{3}}^{'}\}\rightarrow\{H_{1},H_{2},H_{3}\}$. $\{0,0,{H_{3}}^{''}\}$ represents the high temperature vacuum expectation value at the onset of the phase transition while $\{0,{H_{2}}^{'},{H_{3}}^{'}\}$ and $\{H_{1},H_{2},H_{3}\}$ represent the low temperature vacuum expectation values at the second and third steps, respectively. During the initial step, the strength of the phase transition along the $H_{2}$ is a strong one, as indicated by $\zeta_{c,2}=1.16$. Conversely, the transition along the $H_{3}$ direction is weak, with $\zeta_{c,3}=0.001$. These values are observed at the critical temperature $T_{c}=726$ GeV. In the second step of the phase transition, all field directions experience a second-order phase transition. The first-order phase transitions are associated with the formation and subsequent collapse of bubbles of the new phase, which can generate powerful gravitational waves. However, in the second step of the phase transition, a second-order phase transition occurs, the change in the system's properties is gradual, preventing the formation of such bubbles and thus limiting the production of gravitational waves.

\section{Dark matter phenomenology}\label{dark-matter}
In this section, we explore the phenomenological implications of the pNGB dark matter (DM) candidate on the parameter space that provides a strong first-order phase transition. To achieve this goal, we thoroughly study dark matter relic density, and its direct and indirect detection constraints. The latest results on the dark matter relic density as obtained from the Planck satellite's observation indicate \cite{ Planck:2018vyg}
\begin{eqnarray}
    \Omega_{\rm DM}h^2 = 0.120 \pm 0.0012, 
\end{eqnarray}
which leads to $2\sigma$ bound 
\begin{eqnarray}
    0.1176 < \Omega_{\rm DM}h^2 < 0.122,
\end{eqnarray}

\begin{figure}
\includegraphics[scale=0.2]{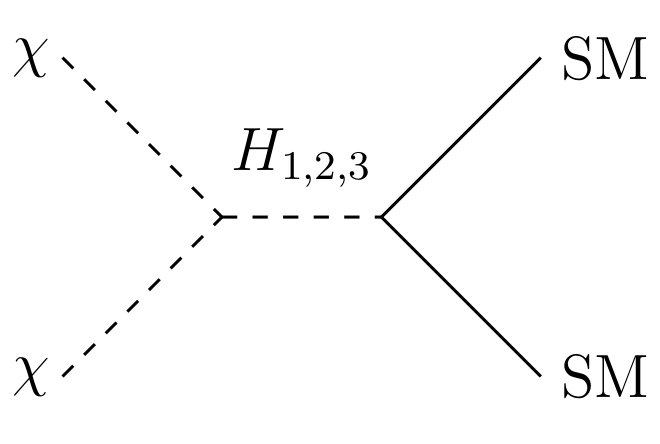}
\includegraphics[scale=0.2]{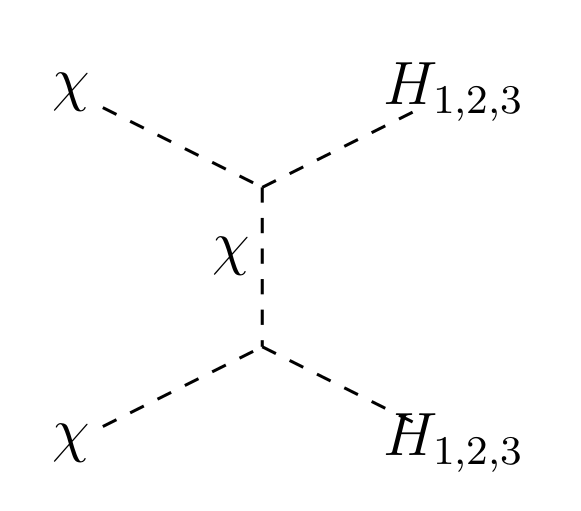}
\includegraphics[scale=0.2]{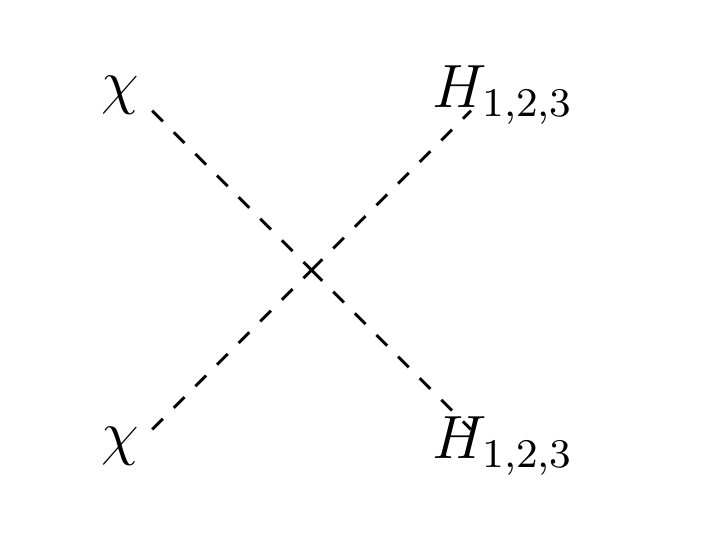}
\caption{\it Some representative Feynman diagrams of DM annihilation processes.}
\label{fig:feyn}
\end{figure}

In our present scenario, the unbroken $\mathcal{Z_{\rm 2}}$ symmetry ensures the stability of the pNGB field $\chi$ over cosmological timescales, making it a suitable DM candidate. The mass of the $\chi$ particle, denoted by $m_\chi$, is determined by the $\mathcal{Z_{\rm 3}}$ breaking scale $\mu_3$ and the vacuum expectation value (vev) $(v_s)$ of the complex scalar field $S$, as shown in Eq.\eqref{mass:chi}. According to the scalar potential $V_0(H, S,\Phi)$ in Eq.\eqref{eq:potential}, $\chi$ can be produced in the early universe's thermal plasma through interactions with the Standard Model Higgs or other singlet scalar fields. As these interaction rates fall below the universe's expansion rate, quantified by the Hubble parameter $H$, the comoving number density $(Y_\chi)$ of $\chi$ decouples from the thermal bath. The present abundance of $\chi$ is primarily influenced by the annihilation processes depicted in Figure \ref{fig:feyn}. Solving the Boltzmann Equation is essential to track the comoving number density of DM.
\begin{equation}\label{yield}
    \frac{dY_{\chi}}{dx}=\frac{\beta (x) s(x)}{H(x) x}\left[\left\langle\sigma_{\left(\chi \chi \rightarrow {\rm SM}\, {\rm SM}\right)}v\right> \left((Y_{\chi}^{eq})^2 - Y^2_\chi\right)\right],
    \end{equation}
    
The various components of Eq.(\ref{yield}) are the following: the co-moving number density $Y_\chi = \frac{n_\chi}{s}$, where $s$ is the entropy density,  $x = m_{\chi}/T$, with $T$ representing the thermal bath's temperature. $Y^{eq}_{\chi}$ represents the equilibrium co-moving density of $\chi $. The term $\left\langle\sigma_{\left(\chi \chi \rightarrow SM, SM\right)}v\right \rangle$ refers to the thermally averaged annihilation cross section of DM into SM particles. Here, $H$ denotes the Hubble parameter. The parameter $\beta$ is expressed as $\beta (T) = \frac{g_{\star}^{1/2}(T) \sqrt{g_{\rho}(T)}}{g_{s}(T)}$, with $g_{\rho}$ and $g_s$ representing the effective degrees of freedom (DOFs) related to the energy density and entropy density, respectively, and $g_{\star}^{1/2}=\frac{g_s}{\sqrt{g_\rho}} \left( 1+ \frac{1}{3} \frac{T}{g_{s}(T)} \frac{dg_s (T)}{dT}\right)$. Some representative Feynman diagrams in Figure \ref{fig:feyn} illustrate that the present relic abundance of the DM in the universe is mainly determined by the parameter $\lambda_{HS}$ and scalar sector mixing angles.
To obtain the relic density, we solve the Boltzmann Equation by using the publicly available package called \textsf{micrOMEGAs} \cite{Belanger:2014vza} where the model information has been provided by using the package \textsf{FeynRules} \cite{Alloul:2013bka}.

\begin{figure}[ht]
    \centering
    \includegraphics[width=12 cm]{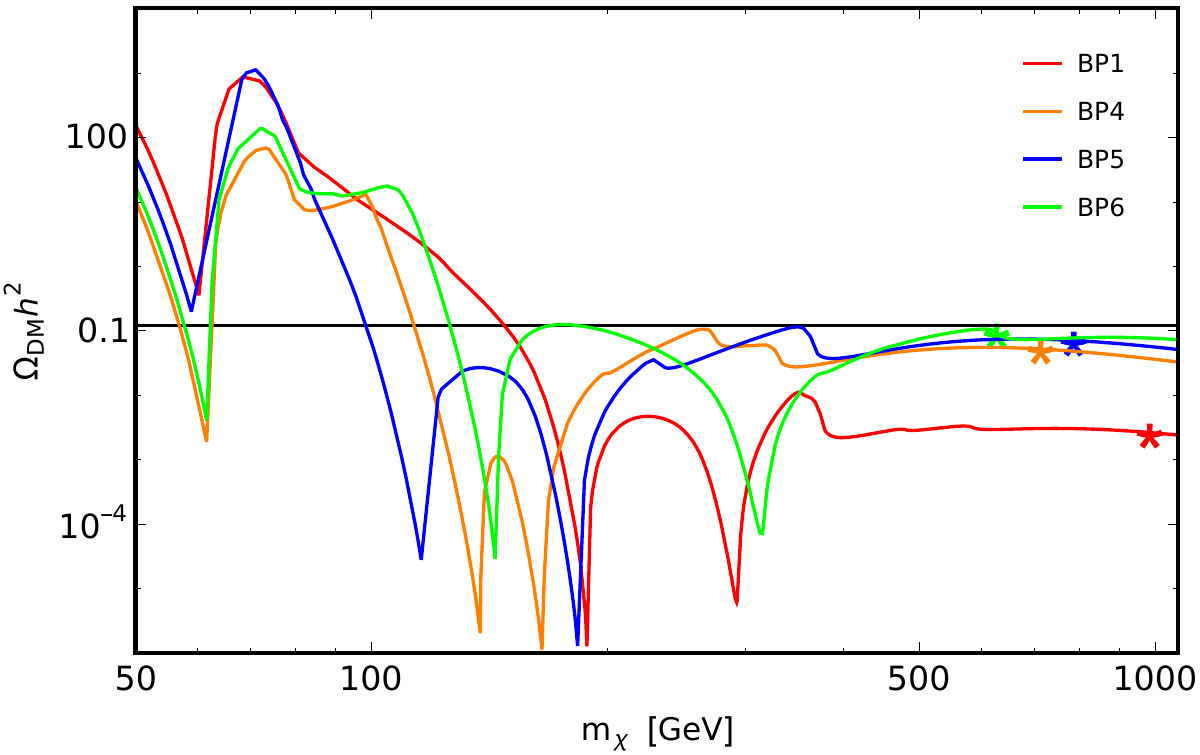}
    \caption {\it The figure shows the variation of dark matter relic density as a function of dark matter mass for the four benchmark points specified in Table \ref{benchmarkpoints}, assuming other parameters remain constant. The stars indicate the dark matter mass values corresponding to the four benchmark points (BP1 in red, BP4 in orange, BP5 in blue, and BP6 in green) as presented in Table \ref{benchmarkpoints}. The horizontal black line corresponds to $\Omega_{\rm DM}h^2 = 0.12$ (observed thermal relic density).} 
    \label{DMrelic}
\end{figure}

In Figure \ref{DMrelic} we show the variation of the relic density with dark matter mass for some representative BPs. The generic feature of relic density is represented in the figure along with various resonance effects. For a given BP, we vary the dark matter mass, while keeping other input parameters fixed as given in Table \ref{benchmarkpoints}. 

The plot illustrates the relic density curves for some chosen benchmark points (BPs), each represented by different colors (see Figure \ref{DMrelic} for details). The chosen benchmark points are such that they satisfy direct detection also as discussed later in Sec.\ref{dds}. The prominent dips at $m_\chi = m_{H_i}/2$ (where $m_\chi$ is the dark matter particle mass and $m_{H_i}$ are the masses of three scalar particles, $H_i$) correspond to $s$-channel annihilation processes mediated by these respective $H_i$ scalars. The starred data points indicate the dark matter masses representing those BPs. Notably, most BPs (BP1, BP2, BP4, BP5, and BP6) show an underabundance of dark matter relic density due to high annihilation cross section. Therefore, to quantify the dark matter abundance we define a ratio : 
\begin{eqnarray}
f_\chi = \frac{\Omega_\chi h^2}{\Omega_{\rm DM}h^2}
\end{eqnarray} 

where we assume that it is not mandatory for the pNGB field $\chi $ to be the only DM component; it could make a subdominant contribution to the total DM density, while still interacting strongly enough with nuclei to be potentially detectable at direct detection experiments. If $f_\chi < 1$, the model predicted relic density is suppressed compared to that of the observed value by $f_\chi$ which corresponds to larger values of scalar sector couplings required for the strong first-order phase transition. 

However, BP3 and BP7 exhibit relic densities closer to the experimentally observed value. Our primary goal is to achieve a strong first-order phase transition for either or both the SM and BSM Higgs fields. This requires high values of the quartic couplings $\lambda_{HS}$ and $\lambda_{S \phi}$, which increase as $v_s$ (the vacuum expectation value of the scalar field $s$) decreases. As shown in Table \ref{benchmarkpoints}, higher values of $v_s$ generally lead to higher relic densities, approaching the observed value. Consequently, $v_s$ is a critical tunable parameter for achieving the desired relic density in our model. Among the BPs in Table \ref{PTSM} (SM case), BP3 has the highest $v_s$ and thus the highest relic density (0.12). Similarly, BP7 in Table \ref{PTBSM} (BSM case) has the highest $v_s$ among its counterparts, with a relic density of 0.119.
~

\subsection{Direct Detection Search}
\label{dds}
\begin{figure}[h]
    \centering
    \includegraphics[width=5 cm]{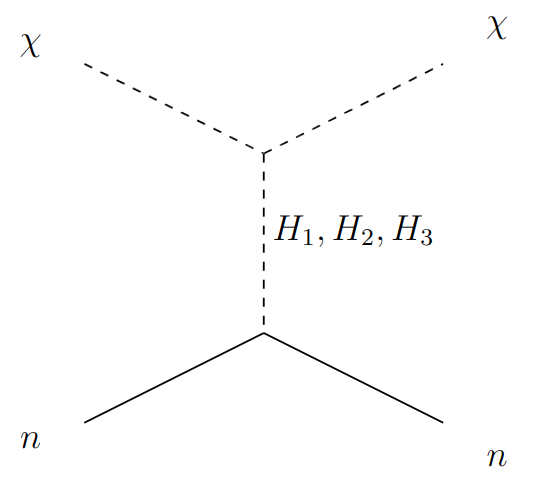}
    \caption{\it Feynman diagrams for spin-independent DM-nucleon scattering process. $n$ is the nucleon and $\chi$ is the DM candidate.}
    \label{ddfeyn}
\end{figure}

Over the years, many studies on pseudo-Nambu-Goldstone boson (pNGB) dark matter have explored various theoretical frameworks and detection prospects. Investigations on direct detection mechanisms for light dark matter was studied by Essig et al. (2012) \cite{PhysRevD.85.076007}. A pNGB is an attractive candidate for dark matter as it can evade the severe limits of constraints in the recent DM direct detection experiments \cite{Abe:2021byq}. In general, pNGB DM of any $\mathcal{Z_{\rm 2}}$ symmetric model has vanishing direct detection cross section at the tree level in the limit $t\rightarrow 0 $ \cite{Gross:2017dan} where $t$ is the momentum transfer of the DM with the nucleon. However, the introduction of any additional symmetry-breaking parameter will give a non-zero direct detection cross section in the limit of zero momentum transfer \cite{Kannike:2019mzk}. In our model, the direct detection cross section directly depends on the $\mathcal{Z_{\rm 3}}$ breaking scale $\mu_{3}$ and also the quartic couplings $\lambda_{HS}$.  If there is more than one symmetry-breaking parameter in a model, some specific choice of those parameters can give zero direct detection cross section at tree level in the limit $t\rightarrow0$ \cite{Alanne:2020jwx}. As our model only has one such symmetry-breaking parameter, we don't have a cancellation mechanism of DM - nucleon interaction at the tree level for any non-zero value of the parameter $\mu_{3}$.  We use the \textsf{micrOMEGAs} \cite{Belanger:2014vza} package to calculate the direct detection cross section of dark matter-nucleon interactions. The relevant Feynman diagrams for the calculation of the DD cross-section are shown in Figure \ref{ddfeyn}. The expression  of spin-independent direct detection differential cross section is given by \cite{Alanne:2020jwx}
\begin{eqnarray}\label{dd1}
    \frac{d \sigma^{\rm SI}}{d\Omega}=\frac{\lambda_{\rm SI}^{2} f_{n}^{2}m_{n}^2}{16 \pi^2 m_{\chi}^2}\left(\frac{m_{\chi}m_{n}}{m_{\chi}+m_{n}}\right)^2
\end{eqnarray}
where the effective Higgs-nucleon coupling $f_{n}=0.3$, the nucleon mass $m_{n}=0.946$ GeV \cite{PhysRevD.85.051503,ALARCON2014342,Cline:2012hg}, and the effective DM-nucleon coupling $\lambda_{\rm SI}$ is given by
\begin{eqnarray}
    \label{dmcoup}
    \lambda_{\rm SI}^{2}=\frac{(4m_{n}^2
-t)}{m_{n}^{2}v_{h}^2}\left[\frac{\lambda_{H_{1}\chi\chi}(c_{\alpha_{1}}c_{\alpha_{2}})}{t-m_{H_{1}}^2}+\frac{\lambda_{H_{2}\chi\chi}(s_{\alpha_{1}}c_{\alpha_{2}})}{t-m_{H_{2}}^2}+\frac{\lambda_{H_{3}\chi\chi}(s_{\alpha_{2}})}{t-m_{H_{3}}^2}\right]^2  
\end{eqnarray}
   where $c_{\alpha_i}=\cos\alpha_i$ and $s_{\alpha_i}=\sin\alpha_i$ and the quantities $\lambda_{H_{i}\chi\chi}$  are given by,
   \begin{align}
   & \lambda_{H_{1}\chi\chi}  = -
   \frac{(m_{H_{1}}^2+m_{\chi}^2)}{v_{s}}(s_{\alpha_{1}}c_{\alpha_{2}}),\nonumber\\
  & \lambda_{H_{2}\chi\chi}  = -
   \frac{(m_{H_{2}}^2+m_{\chi}^2)}{v_{s}}(c_{\alpha_{1}}c_{\alpha_{3}}-s_{\alpha_{1}}s_{\alpha_{2}}s_{\alpha_{3}}),\nonumber\\
  & \lambda_{H_{3}\chi\chi}  = 
   \frac{(m_{H_{3}}^2+m_{\chi}^2)}{v_{s}}(c_{\alpha_{1}}s_{\alpha_{3}}+s_{\alpha_{1}}s_{\alpha_{2}}c_{\alpha_{3}}).
   \label{DDvf}\end{align}
Clearly, we can see that in the limit $t\rightarrow0$, the value of $\lambda_{\rm SI}$ in Eq.(\ref{dmcoup}) does not vanish.

\begin{figure}[ht]
  \subfigure[]{
  \includegraphics[scale=0.34]{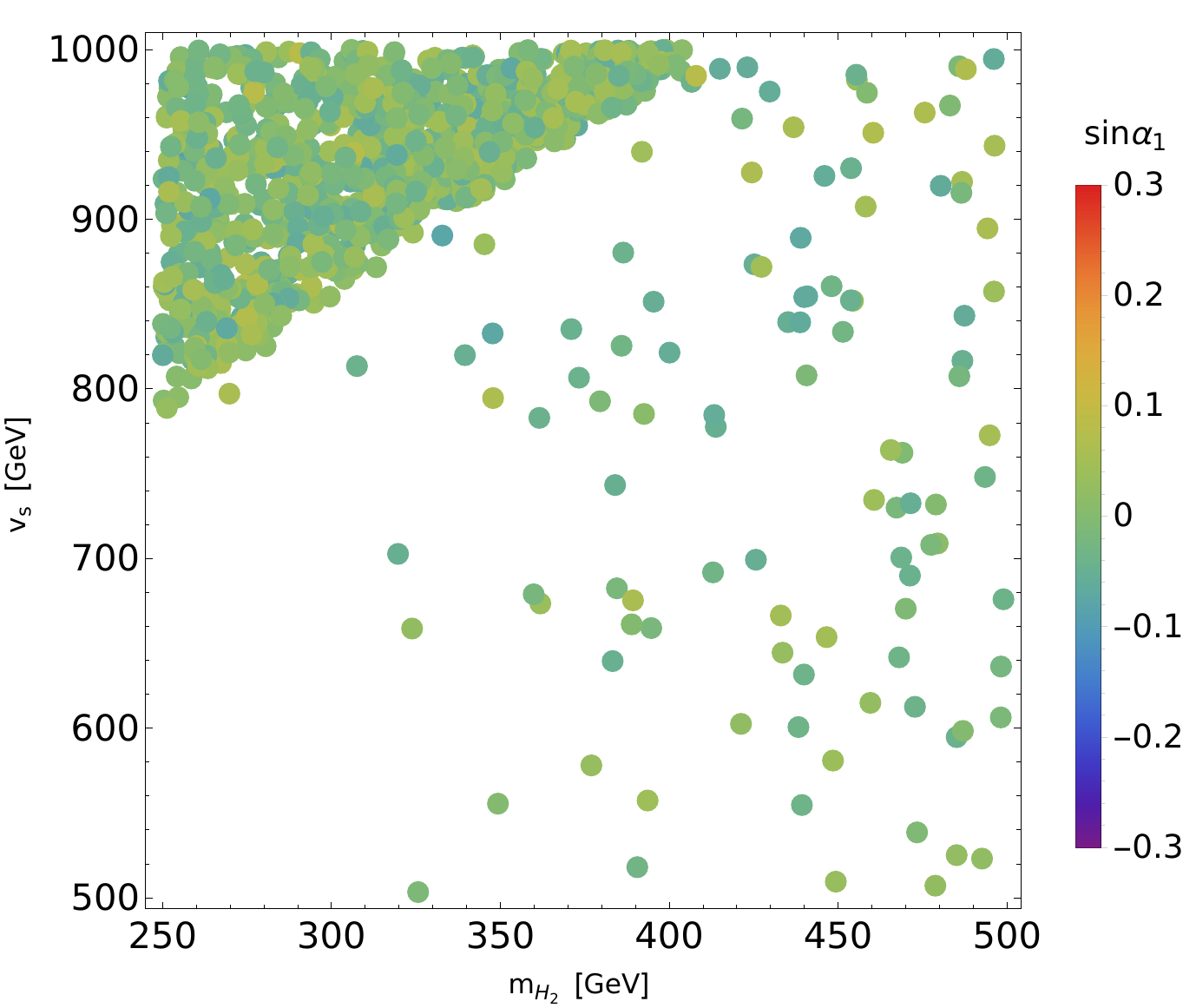}}
  \subfigure[]{
  \includegraphics[scale=0.34]{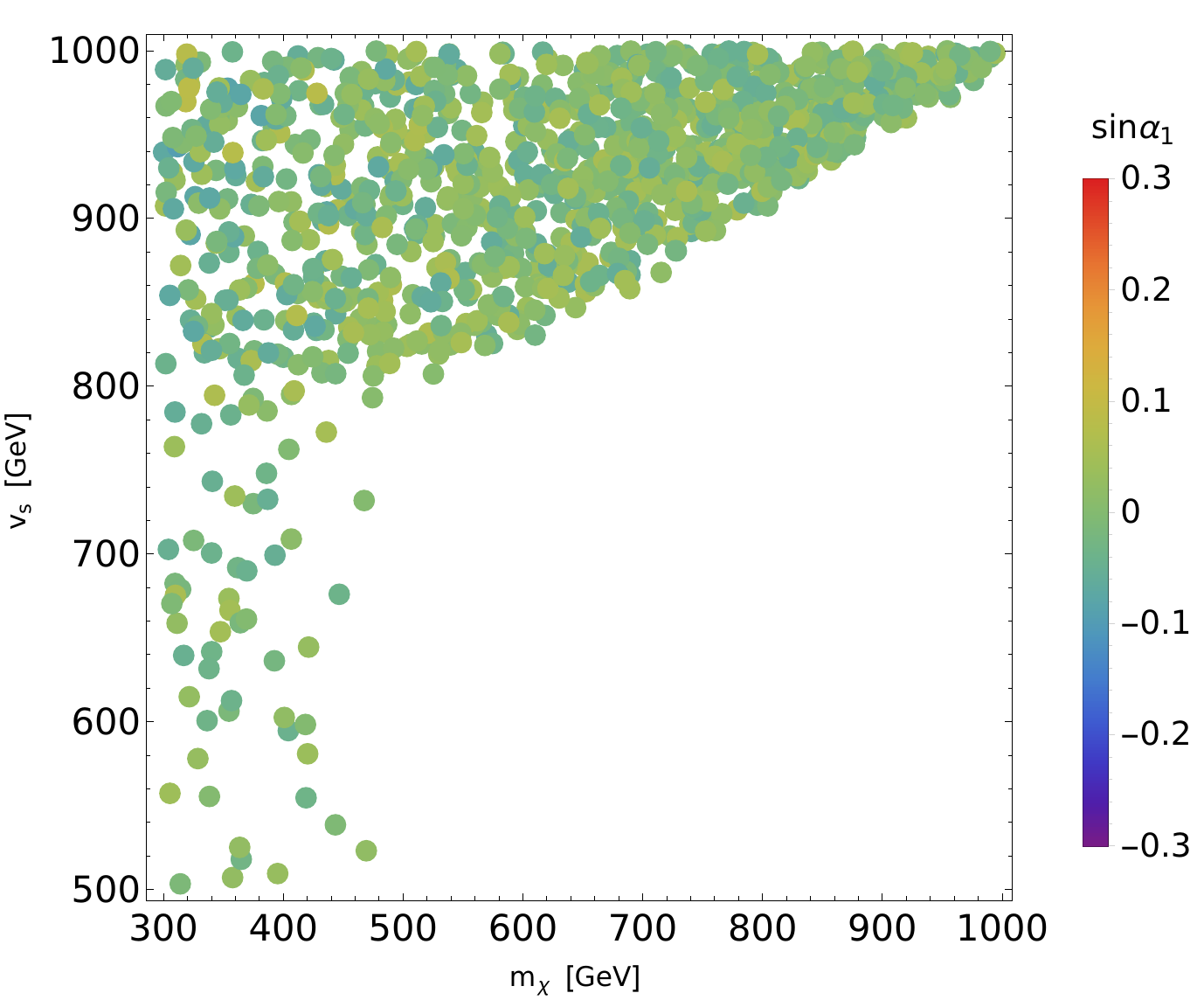}}
   \caption{\it Available parameter space in $m_{H_2}\, -\, v_s$ (a) and $m_\chi\, -\, v_s$ (b) planes. The points satisfy the 2 $\sigma$ range of relic density $\left(0.12 \pm 0.001 \right)$ as well as the latest LZ bound \cite{LZ:2022lsv}.} 
    \label{fig:DM_scan1}
\end{figure}

A significant portion of the parameter space, particularly in the low-mass region, cross-section remains sufficiently small which can evade the strong DD constraints from experiments like XENON1T \cite{XENON:2018voc}, PandaX-II \cite{PhysRevD.85.051503}, LZ \cite{LZ:2022lsv}. This is because lighter $m_{\chi}$ corresponds to a smaller value of $\mu_3$ hence the smaller cross-section. It is customary to show the variation of relic density and direct detection cross-section with respect to relevant model parameters. For this, we perform a parameter scan over the different suitable parameters of the model. We see that parameters like $\sin\alpha_{2}$, $\sin\alpha_{3}$ and $v_{\phi}$ do not affect the DM relic density significantly. This promptly simplifies our parameter scan strategy by setting their values at  $\sin\alpha_{2} = \sin\alpha_{3} =0.1$, and $v_{\phi}$=500 GeV, during the scan process.

\begin{figure}[hbt!]
   \subfigure[]{
  \includegraphics[scale=0.32]{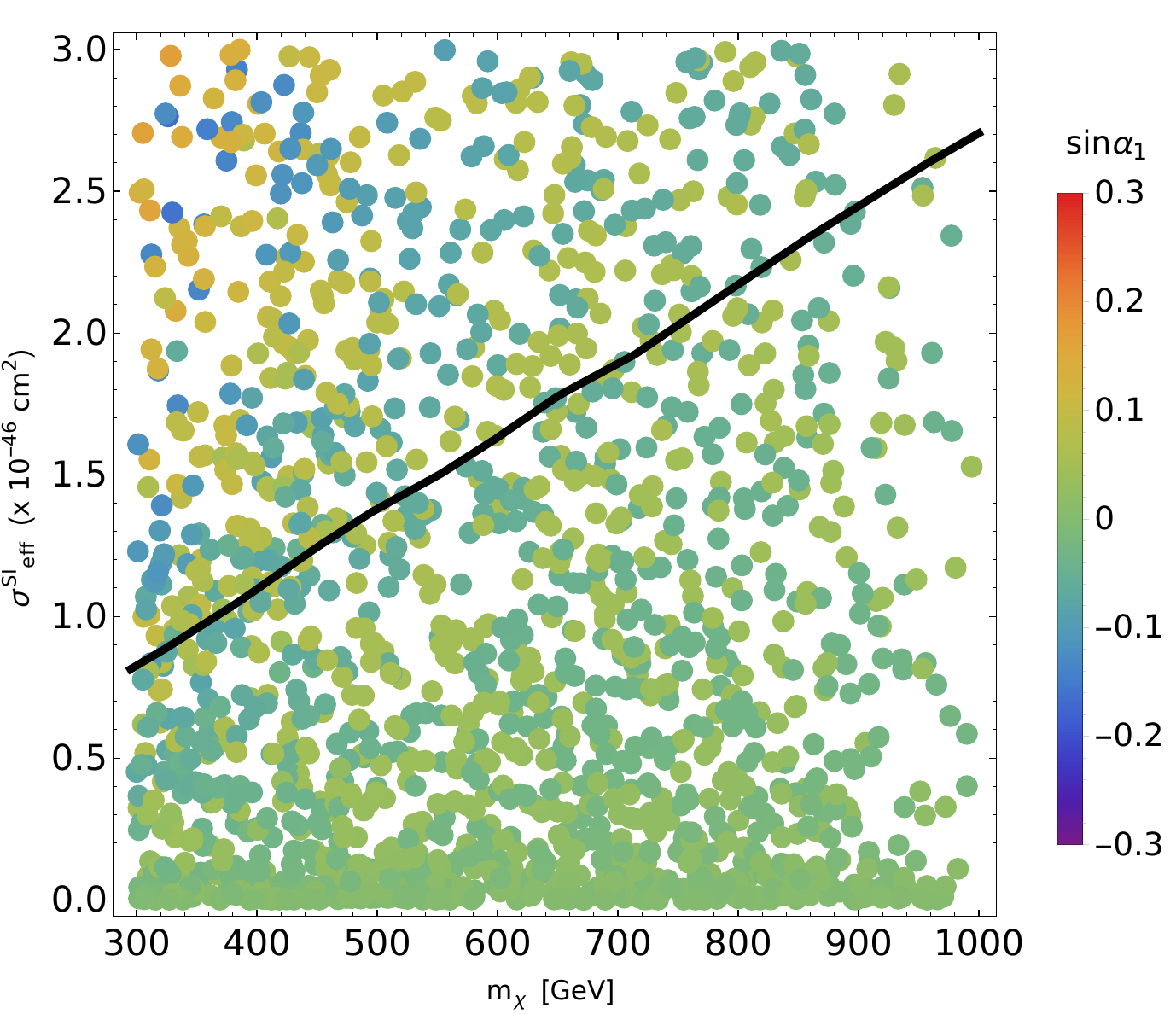}
  \label{fig:DD1}}
  \subfigure[]{
  \includegraphics[scale=0.32]{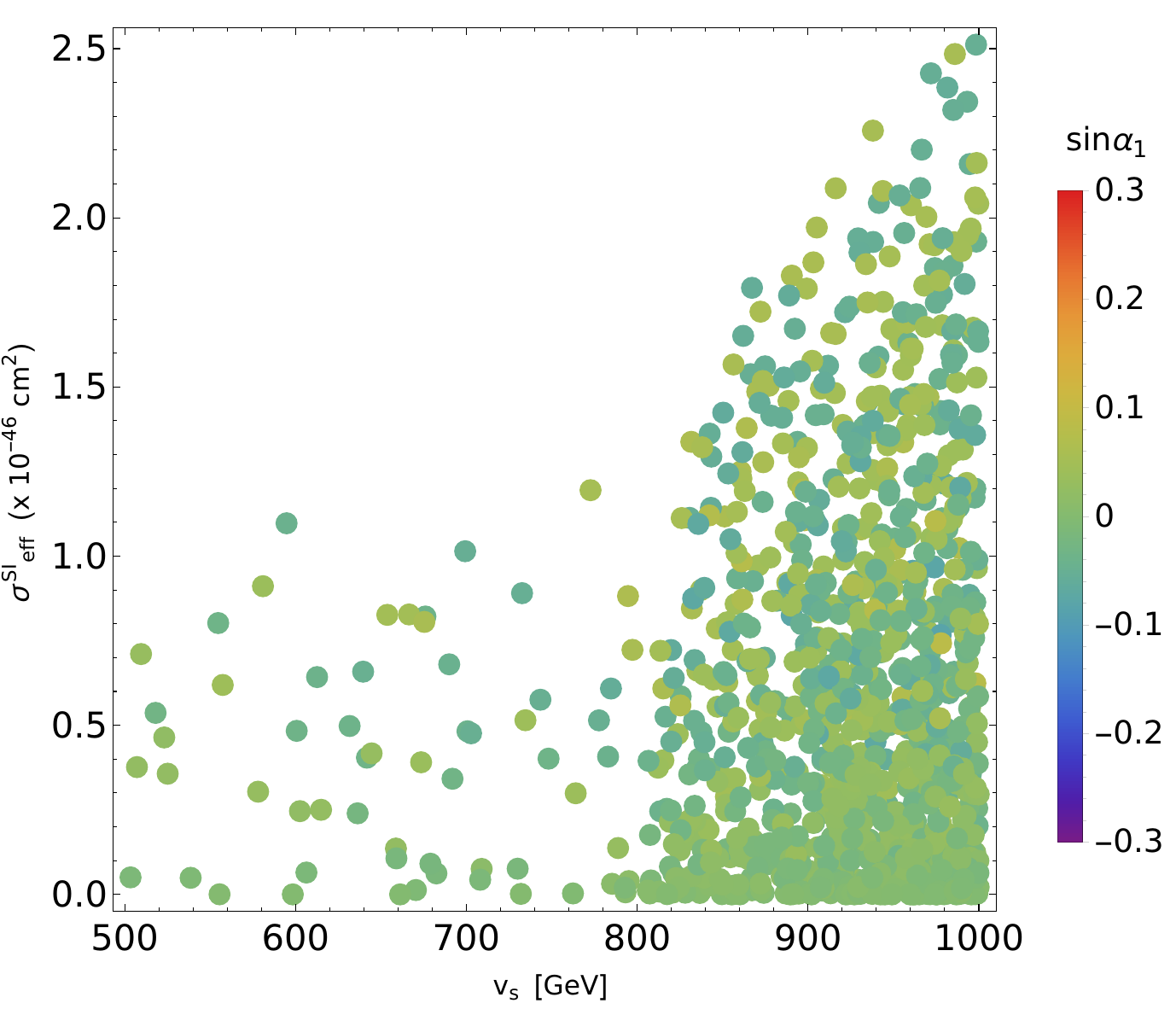}
      \label{fig:DD2}}
    \caption{\it Scatter plot of points which satisfy the relic density within 2$\sigma$ range of 0.12 in the plane of $m_{\chi}$ and $v_{s}$ with $\sigma^{\rm SI}_{\rm eff}$ are shown in Figure (a) and (b) respectively. The parameter scan is the same for generating Figure \ref{fig:DM_scan1}. The bold black line in Figure  (a) corresponds to the latest LZ bound~\cite{LZ:2022lsv}. Only the points below the curve are allowed from Direct Detection Searches. In Figure  (b) all the points are allowed from Direct Detection searches.  }
\end{figure}

The parameters that play a crucial role and significantly affect the relic density are the DM mass $m_{\chi}$, the vev of the complex scalar $v_s$, the masses of the real singlet scalars $m_{H_2}$, $m_{H_3}$ and the mixing angle $\sin{\alpha_1}$. Finally, we show interesting correlations among different parameters in Figure \ref{fig:DM_scan1}. The colour bar represents the variation of the mixing angle $\sin{\alpha_1}$. To achieve a relic density around 0.12, the choice of a high value of $v_s$ is preferable as mentioned earlier. Eq.(\ref{qcoupling}) reveals that the parameters $\lambda_{HS}$ and $\lambda_{S\Phi}$ feature $v_{s}$ in the denominator suggesting that this quantity increases with relic density due to its inverse relationship with annihilation cross-section which makes $v_{s}$  the most adjustable parameter for achieving the desired relic density by changing its value. As mentioned earlier, direct detection cross-section depends on the $\mathcal{Z}_3$ breaking scale $\mu_3 $ and the quartic coupling $\lambda_{HS}$. However, in our analysis, we express them as functions of the free parameters like $m_\chi$ and $v_s$. The spin-independent direct detection cross section $\sigma^{\rm SI } $ obtained from \textsf{micrOMEGAs} has been scaled by $f_{\chi}$ to compare with the upper bounds of experiments~\cite{Cline:2012hg}: 
 \begin{eqnarray}
     \sigma^{\rm SI}_{\rm eff}= f_{\chi} {\sigma}^{\rm SI},~~(f_\chi \leq 1)
     \label{fchi}
 \end{eqnarray} 
 
The direct detection cross-section (effective) is shown as a function of $m_\chi$ and $v_s$ in Figure \ref{fig:DD1} and Figure \ref{fig:DD2} respectively. From Figure \ref{fig:DD1}, we see that to satisfy the latest experimental results of DD cross section, a very small value of $\sin\alpha_{1}$ is preferred for a wide range of dark matter mass and we impose the most recent bound from the LZ~\cite{LZ:2022lsv}. From Figure \ref{fig:DD2}, we see that a very small value of $\sin\alpha_{1}$ along with a high value of $v_{s}$ is preferred to satisfy both the correct relic density (within $2\sigma$) and the direct detection cross section bound. We present the values of the observables relic density and spin-independent direct detection cross section in Table \ref{DMoutput}. It is seen that BP2, BP3, and BP7 are excluded from the latest LZ results~\cite{LZ:2022lsv}. Thus, we only have 4 Benchmark Points that survived after applying direct detection constraints. Further, we will proceed with these BPs only.

\begin{table}[h]
  \resizebox{ \textwidth}{!}{
    \begin{tabular}{|c|c|c|c|c|c|c|c|}
      \hline Observables&  BP1 & BP2 & BP3 & BP4&BP5 &BP6 &BP7    \\\hline \hline
    
    $\Omega_{\rm DM}h^2$&0.00261&0.095&0.12&0.0517&0.071&0.0913&0.119\\\hline
    ${\sigma^{\rm SI}_{\rm eff}}$ $\rm (cm^2)$&$6.11\times10^{-51}$&$1.9\times10^{-46}$&$6.78\times10^{-46}$&$3.76\times10^{-47}$&$1.77\times10^{-47}$&$2.46\times10^{-47}$&$1.125\times10^{-45}$\\\hline
    \end{tabular}
 }   
    \caption{\it Relic Density and Direct Detection cross section values for all the Benchmark points.}
    \label{DMoutput}
    
\end{table}

\subsection{Indirect Detection Search}
 Various ground and space-based experiments look for the indirect signature of dark matter \cite{Fermi-LAT:2013thd,Fermi-LAT:2015qzw},  where they search for the signal of dark matter pair annihilation or decay in the galactic halo or outside it. Such experiments assume the existence of DM (WIMPs) in sufficient densities within certain astrophysical environments—such as the Sun, Earth, Dwarf Spheroidal Galaxies (dSph), or the Galactic Center. DM particles pair annihilate or decay in these astrophysical objects, thereby creating SM particles that undergo decay, showering, and hadronization. These long-chain processes result in the fluxes of stable SM particles such  as gamma rays, neutrinos, and various light-charged particles. All these can be searched in cosmic rays as a signal of DM by ground-based telescopes such as H.E.S.S.\cite{PhysRevLett.117.111301} and CTAs \cite{Hofmann:2023fsn} or satellites such as AMS \cite{PhysRevLett.113.221102}, PAMELA \cite{PAMELA:2010kea}, and Fermi-LAT \cite{Fermi-LAT:2015att}. If DMs are produced thermally in the early universe, then currently the velocity averaged DM self-annihilation cross-section has a natural value of $\langle\sigma_{\rm ann} v \rangle \sim 3\times 10^{-26}~\frac{{\rm cm}^3}{\rm s}$. The Fermi-LAT looks for gamma-ray signals that might be produced by the annihilation of dark matter particles in dSphs, which are faint galaxies with high dark matter densities compared to their visible matter. Meanwhile, in the absence of a discernible $\gamma $ ray signal from dSphs can impose  $95\% $ C.L. upper limits on the $\gamma$-ray flux from the target. Such upper limits on the $\gamma $ ray fluxes from DM annihilation can be translated to constraints in the two-dimensional plane of the DM mass and the thermally averaged DM pair-annihilation cross section
 $\langle\sigma v\rangle$ \cite{Fermi-LAT:2013thd}.
\setlength{\tabcolsep}{11pt}
\begin{table}[]
    \centering
    \begin{tabular}{| c | c | c | c | c |} \hline
         BP&1&4&5&6  \\\hline \hline
         $m_{\chi}$ (GeV)&985.94&714.46&787.2&628.44\\\hline
         $ {\langle{\sigma v} \rangle}_{\rm eff} (\frac{cm^3}{s}) $ &$5.72\times10^{-28}$&$1.02\times10^{-26}$&$1.4\times10^{-26}$&$1.42\times10^{-26}$\\\hline
         Dominant   &$H_2 H_2$ (84.2\%)&$H_3 H_3$ (68.8\%)&$H_3 H_3$ (88.8\%)&$H_2 H_2$ (54.2\%)\\Annihilation&$H_2 H_3$ (8.6\%)&$H_2 H_3$ (16.2\%)&$H_2 H_2$ (5.4\%)&$H_1 H_3$ (17.2\%)\\channels&$H_3 H_3$ (6.9\%)&$H_2 H_2$ (13.3\%)&$H_2 H_3$ (4.3\%)&$W^{+} W^{-}$ (11.7\%)\\\hline
         $H_{2}$   &$W^{+}W^{-}$ (51\%)&$W^{+}W^{-}$ (58\%)&$W^{+}W^{-}$ (70.6\%)&$W^{+}W^{-}$ (63.3\%)\\decay&$Z Z$ (23.5\%)&$Z Z$ (25.1\%)&$Z Z$ (29.1\%)&$Z Z$ (27.8\%)\\modes&$H_{1} H_{1}$ (19.7\%)&$H_{1} H_{1}$ (16.7\%)&$b \Bar{b}$ (3.3\%)&$H_{1} H_{1}$ (8.8\%)\\\hline
         $H_{3}$   &$W^{+}W^{-}$ (45\%)&$W^{+}W^{-}$ (53.7\%)&$W^{+}W^{-}$ (37.4\%)&$W^{+}W^{-}$ (46.9\%)\\decay&$Z Z$ (21.7\%)&$Z Z$ (24.2\%)&$Z Z$ (17.2\%)&$Z Z$ (22.8\%)\\modes&$H_{1} H_{1}$ (13.9\%)&$H_{1} H_{1}$ (22\%)&$H_{1} H_{2}$ (25.6\%)&$H_{1} H_{1}$ (14.3\%)\\\hline
    \end{tabular}
    \caption{\it Total annihilation cross-section of DM and relative contribution from different final states are shown.}
    \label{tabindir}
\end{table}

 \begin{figure}[h]
    \centering
    \includegraphics[width=12 cm]{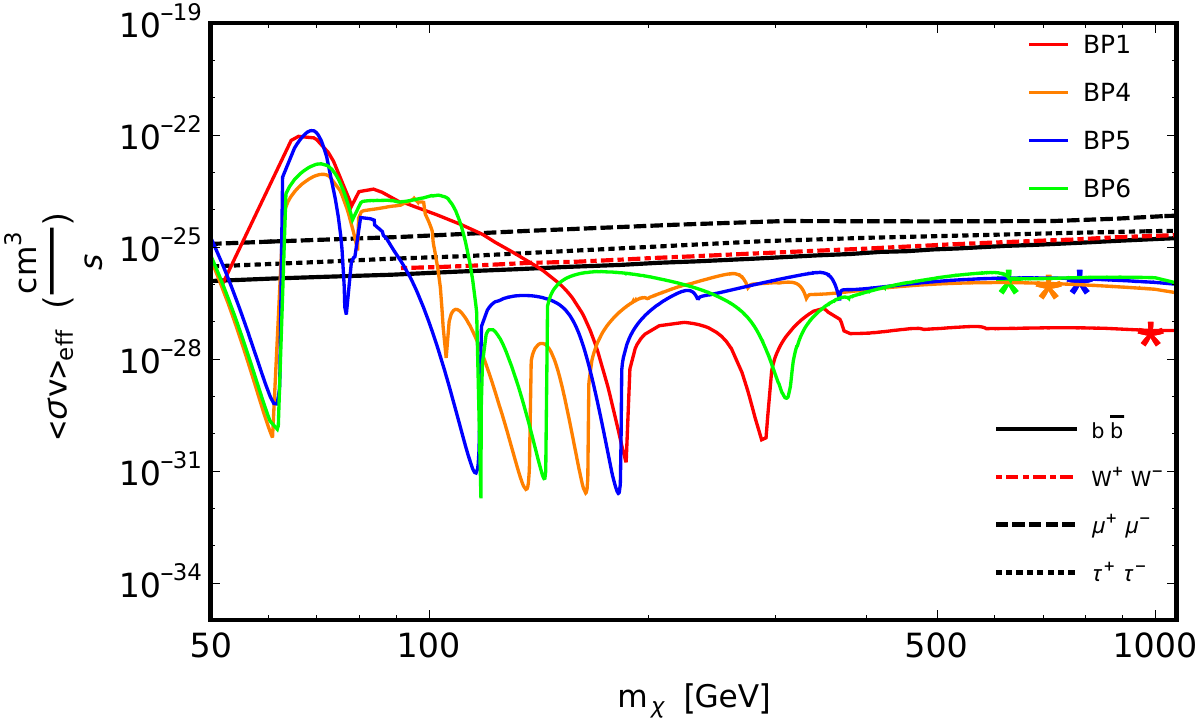}
    \caption {\it The figure shows the variation of the effective dark matter annihilation cross-section as a function of dark matter mass for the four benchmark points specified in Table \ref{tabindir}, assuming other parameters remain constant. The stars indicate the dark matter mass values corresponding to the four benchmark points (BP1 in red, BP4 in orange, BP5 in blue, and BP6 in green) as presented in Table \ref{tabindir}.}
      \label{DMindirect}
\end{figure}

In our model, the dark matter being a pNGB particle annihilates via four-point contact interactions and through heavy scalar-mediated $s$-channel processes. Depending on the mass of the DM, the SM gauge bosons $W^+W^-$ dominate the final state, followed by $ZZ$ and $H_1H_1$. In Table \ref{DMoutput}, we show four representative points BP1, BP4, BP5, and BP6 with DM mass, which satisfy the relic density constraints as well as the direct and indirect detection limits. For these benchmark points, we first estimate the thermally averaged annihilation cross section $\langle \sigma v \rangle $ using \textsf{micrOMEGAs} and properly scale it with $f_{\chi}^2$ to get the effective annihilation cross-section $\langle \sigma v \rangle _{\rm eff} (= \langle \sigma v \rangle \times f^2_\chi )$ to compare with experimental upper bounds from Fermi-LAT and MAGIC focused on dwarf spheroidal galaxies \cite{MAGIC:2016xys,Hess:2021cdp}. In Figure~\ref{DMindirect}, we show the variation of $\langle \sigma v \rangle_{\rm eff} $ as a function of dark matter mass. Here, the red, orange, blue, and green curves represent our model predicted annihilation cross-section corresponding to benchmark points BP1, BP4, BP5, and BP6 respectively. It is worth mentioning here that for a given benchmark point while keeping all other parameters fixed, we vary the dark matter mass to show the general behavior of the effective annihilation cross-section with dark matter mass. The symbol stars represent the actual value of the dark matter mass for the chosen benchmark point. A resonance enhancement of the dark matter annihilation cross-section occurs when the dark matter mass $(m_\chi)$ is equal to half of the mass of a scalar portal particle $(m_{H_1}, m_{H_2}, m_{H_3})$. This feature is evident at specific dark matter mass configurations (62.5, 188.57, 268.19) GeV for BP1, (62.5, 137.82, 165.82) GeV for BP4, (62.5, 117.47, 183.9) GeV for BP5, and (62.5, 145.04, 315.94) GeV for BP6. Each value in the parenthesis represents ($\frac{m_{H_{1}}}{2}\, \frac{m_{H_{2}}}{2}, \frac{m_{H_{3}}}{2}$). In Figure \ref{DMindirect} the observed dips in the $\langle \sigma v \rangle_{\rm eff}$ at those values of $m_\chi$ are attributed to the scale factor $f_\chi$, which is directly proportional to the relic abundance of the dark matter $\Omega_\chi h^2$ as defined in Eq.(\ref{fchi}). Consequently, at resonance, the $\langle \sigma v \rangle_{\rm eff}$ decreases. The combined results from the Fermi-LAT and MAGIC collaborations \cite{2016} for various predicted annihilation cross-sections are also depicted in Figure \ref{DMindirect}. These constraints provide valuable insights into the potential parameter space for dark matter models.

\section{Gravitational Wave Spectrum} \label{Gravitywave}

A first-order phase transition in the early universe can serve as a primordial source of gravitational wave (GW) generation. The basic characteristic of these waves are stochastic and there are three primary mechanisms for their production as detailed in Caprini et al. \cite{MAGIC:2016xys}:
\begin{enumerate}
\item Bubble Collisions: The energetic collisions of expanding true vacuum bubbles during the phase transition directly contribute to the gravitational wave spectrum.
\item Sound Waves: The propagation of bubbles through the primordial plasma generates acoustic waves. These waves interfere constructively and destructively, 
leading to the formation of shear stress and subsequent gravitational wave emission. 
\item Magnetohydrodynamic Turbulence: Following the collisions of bubble walls, highly dynamic magnetohydrodynamic turbulence develops in the plasma, which acts as 
an additional source of gravitational radiation.
\end{enumerate} 

Therefore, by combining these three possible sources of gravitational wave generation, the total GW energy spectrum is expressed as \cite{Ellis:2020awk}:
\begin{eqnarray}\label{GWtot}     
    \Omega_{\rm GW} h^2 \simeq \Omega_{\rm col} h^2+ \Omega_{\rm sw} h^2+ \Omega_{\rm turb}h^2
\end{eqnarray}

The GW Spectrum is primarily shaped by four key parameters
\begin{enumerate}
\item Latent heat parameter $(\alpha)$: It quantifies the strength of the phase transition and is directly proportional to the latent heat released during the process.
\item Inverse duration parameter $(\beta)$: Determines the characteristic timescale of the phase transition, inversely proportional to its duration. 
\item Nucleation temperature $(T_n)$: The temperature threshold at which the rate of bubble formation within the new phase becomes sufficiently high to enable subsequent bubble growth and coalescence. $T_n$ is lower than $T_c$ and is defined in Eq. (\ref{nuctemp}).
\item Bubble wall velocity $(v_w)$: Represents the speed of the bubble wall, assumed to be relativistic for the strong first-order phase transition. 
\end{enumerate}

A first-order phase transition is characterized by the nucleation and expansion of a spherical bubble. The interior of this bubble follows the equation of state of the broken phase. While the surrounding area remains in the symmetric phase separated by a negligibly thin bubble wall. The difference in free energy between the inside and outside of the bubble generates an effective pressure that drives the bubble's expansion. The parameter $\alpha$ quantifies the strength of the phase transition and is defined as \cite{Giese:2020rtr,Giese:2020znk,Ramsey-Musolf:2024ykk}: 
\begin{equation}
    \alpha = \frac{{\rm D} \theta}{3\omega}
    \label{alphacorr}
\end{equation}
where $\theta$ is the trace of the energy-momentum tensor of the fluid (plasma) and is given by $\rho-3 p$ with $\rho$ and $p$ denoting the energy density and pressure respectively. The term ${\rm D}\theta$ is the difference of the trace in false ($\theta_{f}$) and true ($\theta_{t}$) vacuum respectively. $\omega$ is the enthalpy density in the false vacuum, and all the parameters in Eq.(\ref{alphacorr}) are evaluated at the nucleation temperature $T_{n}$. For the first step of the phase transition, in the false $(f)$ and true vacuum $(t)$, the thermodynamical quantities~\cite{Tian:2024ysd}: pressure ($p$), energy density ($\rho$), and enthalpy density ($\omega$) are expressed as: $p_{f}=\frac{1}{3}a T^{4}$, $\rho_{f}=a T^{4}$, $\omega_{f}=\frac{4}{3}a T^{4}$, $p_{t}= - V(H_{i},T)$, $\rho_{t}= -T\frac{dV(H_{i},T)}{dT} + V(H_{i},T)$, $\omega_{t}=T\frac{dV(H_{i},T)}{dT}$ with $a=\frac{\pi^{2} g_{*}}{30}$ respectively. Putting all the relevant expressions in the Eq. (\ref{alphacorr}), we get the expression of $\alpha$ in terms of the effective potential $V(H_{i},T)$ given by,

\begin{equation}
    \alpha = \frac{\frac{T}{4}\frac{dV(H_{i},T) }{dT} - V(H_{i},T) }{\frac{\pi^{2} g_{*}T^{4}}{30}}\bigg|_{T=T_{n}}.
    \label{alphacorr2}
\end{equation}

The parameter $\frac{\beta}{H_{n}}$ represents the ratio of the inverse of the time taken for the phase transition to complete to the Hubble parameter value at $T_{n}$. It can be expressed as \cite{Nicolis:2003tg}:
\begin{eqnarray}
    \frac{\beta}{H_{n}}= T_{n}\frac{d(S_{3}/T)}{dT}\bigg|_{T=T_{n}}
\end{eqnarray}

For $\beta/H_n \gg $ 1\footnote{A large $\beta/H_n $ allows us to neglect the Hubble expansion of the Universe during the phase-transition process.}, the thin wall and envelope approximation effectively describes the situation. In this model, gravitational waves are predominantly produced by the envelopes, while the overlapping regions are ignored. The contribution of bubble collisions to the GW spectrum, red-shifted up to today is then given by, \cite{Jinno:2016vai}
\begin{eqnarray}
    \Omega_{\rm col} h^2 = 1.67\times10^{-5}\left(\frac{\beta}{H_{n}}\right)^{-2}\left(\frac{\kappa_{\rm col}  \alpha}{1+\alpha}\right)^{2}\left(\frac{100}{g_{*}}\right)^{\frac{1}{3}}\left(\frac{0.11 v^{3}}{0.42 + v^{2}}\right)\left(\frac{3.8\left(\frac{f}{f_{\rm col}}\right)^{2.8}}{1+2.8\left(\frac{f}{f_{\rm col}}\right)^{3.8}}\right)
    \label{omegacol}
\end{eqnarray}
where the peak frequency of bubble collisions is given by, \cite{Jinno:2016vai}
\begin{equation}
    f_{\rm col}=1.65\times10^{-5}\left(\frac{0.62}{1.8-0.1 v_{w}+v_{w}^{2}}\right)\left(\frac{\beta}{H_{n}}\right)\left(\frac{T_{n}}{100}\right)\left(\frac{g_{*}}{100}\right)^{1/6}
    \label{fcol}
\end{equation}
and $\kappa_{\rm col}$ is the efficiency factor of bubble collision and expressed by \cite{Borah:2023zsb},
\begin{equation}
    \kappa_{\rm col}=\frac{0.715 \alpha +\frac{4}{27} \sqrt{\frac{3 \alpha }{2}}}{0.715 \alpha +1}
\end{equation}
Sound waves are generated as the bubble propagates and collides with each other in the plasma. The contribution of sound waves to the GW spectrum is given by, \cite{Hindmarsh:2013xza,Hindmarsh:2016lnk,Hindmarsh:2017gnf,Guo:2020grp}
\begin{equation}
     \Omega_{\rm sw} h^2 =2.65\times10^{-6} \Gamma_{\rm sw}\left(\frac{\beta}{H_{n}}\right)^{-1}v_{w}\left(\frac{\kappa_{\rm sw}  \alpha}{1+\alpha}\right)^{2}\left(\frac{g_{*}}{100}\right)^{\frac{1}{3}}\left(\frac{f}{f_{\rm sw}}\right)^{3}\left(\frac{4}{7}+\frac{3}{7}\left(\frac{f}{f_{\rm sw}}\right)^{2}\right)^{-\frac{7}{2}}
     \label{omegasw}
\end{equation}
The $\Gamma_{\rm sw}$ factor in the above expression appears due to the finite lifetime of the sound waves thus generated, which suppresses the amplitude of the GW spectrum and is expressed as \cite{Hindmarsh:2017gnf},
\begin{equation}
    \Gamma_{\rm sw}=1-\frac{1}{\sqrt{1+2\tau_{\rm sw}H_{n}}}
\end{equation}where $\tau_{\rm sw}$ is the lifetime of the sound waves and is expressed by \cite{Hindmarsh:2017gnf}, 
\begin{equation}
    \tau_{\rm sw}=\frac{(8\pi)^{\frac{1}{3}}}{\beta U}
\end{equation}
 $U$ is the root-mean-squared velocity of the fluid and is expressed as \cite{Hindmarsh:2017gnf},

\begin{equation}
  U=\sqrt{\frac{3}{4} \alpha  \kappa_{\rm sw}}  
\end{equation}
The peak frequency of sound waves contribution is given by, \cite{Hindmarsh:2017gnf}
\begin{equation}
    f_{sw}=1.9\times10^{-5}\left(\frac{1}{ v_{w}}\right)\left(\frac{\beta}{H_{n}}\right)\left(\frac{T_{n}}{100}\right)\left(\frac{g_{*}}{100}\right)^{1/6}.
    \label{fsw}
\end{equation}
The efficiency factor corresponding to the latent heat of the sound waves converting to the motion of plasma is given by \cite{Borah:2023zsb},
\begin{equation}
    \kappa_{\rm sw}=\frac{\alpha }{\alpha +0.083 \sqrt{\alpha }+0.73}
\end{equation}
The plasma is ionized and gives rise to turbulence which is magnetohydrodynamic in nature. This forms another source of gravitational waves. The contribution of magnetohydrodynamic turbulence to the GW spectrum is given by, \cite{Caprini:2009yp}
\begin{equation}
    \Omega_{\rm turb} h^2 =3.35\times10^{-4} \left(\frac{\beta}{H_{n}}\right)^{-1}v_{w}\left(\frac{\kappa_{\rm turb}  \alpha}{1+\alpha}\right)^{\frac{3}{2}}\left(\frac{100}{g_{*}}\right)^{\frac{1}{3}}\left(\frac{\left(\frac{f}{f_{\rm turb}}\right)^{3}}{\left(1+\left(\frac{f}{f_{\rm turb}}\right)^{\frac{11}{3}}\right)\left(1+\frac{8\pi f}{h_{*}}\right)}\right)
    \label{omegaturb}
\end{equation}
where the expression of $h_{*}$ is the  inverse Hubble time during the production of gravitational waves and is given by,

\begin{equation}
    h_{*}= 16.5\times\left(\frac{T_{n}}{100}\right)\left(\frac{g_{*}}{100}\right)^{1/6}.
\end{equation}

The peak frequency of magnetohydrodynamic turbulence contribution is given by, \cite{Caprini:2009yp}
\begin{equation}
    f_{\rm turb}=2.7\times10^{-5}\left(\frac{1}{ v_{w}}\right)\left(\frac{\beta}{H_{n}}\right)\left(\frac{T_{n}}{100}\right)\left(\frac{g_{*}}{100}\right)^{1/6}
    \label{fturb}
\end{equation}

 $\kappa_{\rm turb}$ is the efficiency factor corresponding to the MHD turbulence in the plasma and is given by $\kappa_{\rm turb}= 0.1 \kappa_{\rm sw}$ where we have considered $10\%$  of the bulk motion to be turbulent as suggested by simulations \cite{Borah:2023zsb}.

\begin{figure}[h]
    \centering
    \includegraphics[width=11 cm]{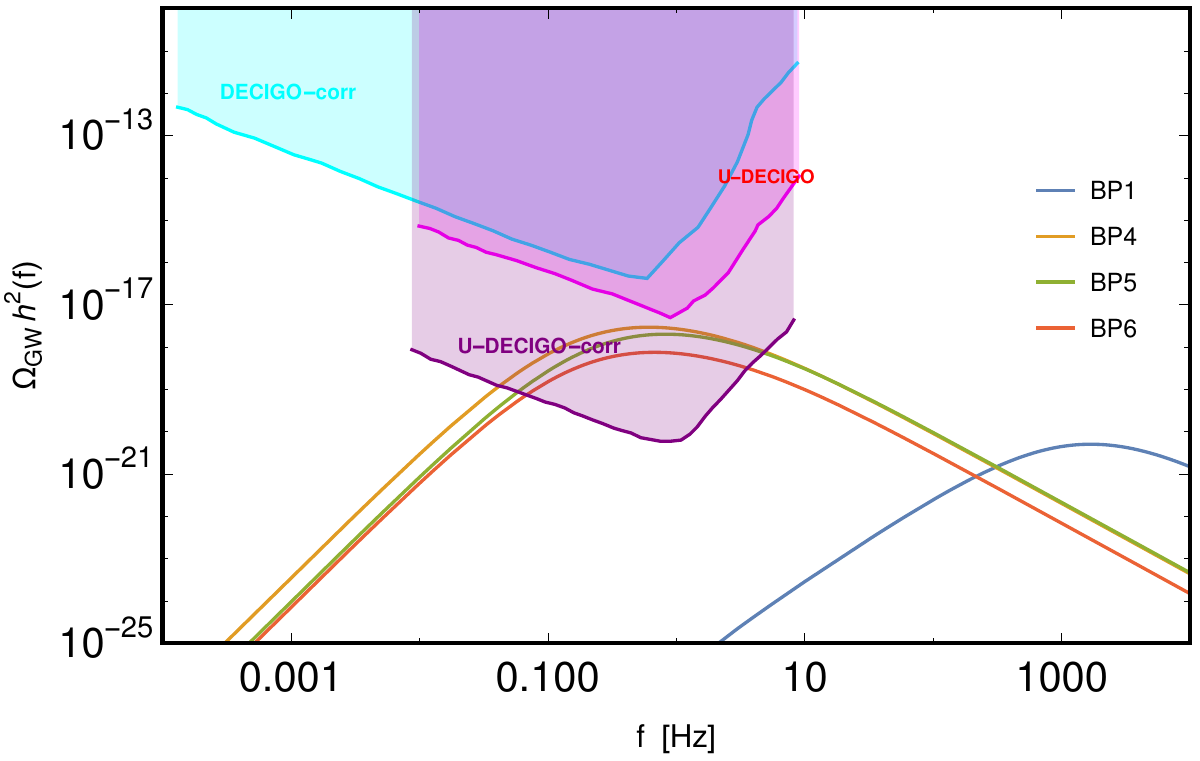}

\caption{\it {This figure presents the gravitational wave (GW) spectrum associated with benchmark points BP1, BP4, BP5, and BP6. We emphasize the sensitivity regions of only those detectors that exhibit a detectable GW spectrum for at least one of the benchmark points. The sensitivity curves for DECIGO, U-DECIGO, and their corresponding correlation analyses are obtained from \cite{Nakayama:2009ce}.}}
    \label{GWplot}
\end{figure}

It can be seen from Eq.(\ref{fcol}), Eq.(\ref{fsw}) and Eq.(\ref{fturb}), the peak frequency (without redshift) of the gravitational wave spectrum depends only on the value of $\frac{\beta}{H_{n}}$ and is independent of $\alpha$.
 
The generated gravitational wave would be possibly detected by the late time observatories such as LISA \cite{Bartolo:2016ami}, Taiji \cite{Ruan:2018tsw}, TianQin \cite{TianQin:2015yph}, Big Bang Observer (BBO)\cite{Cutler:2009qv}, DECi-hertz Interferometer GW Observatory (DECIGO) and Ultimate-DECIGO \cite{Kawamura:2006up}. To determine the detectability of a signal against the background, the most frequently used measure is the Signal-to-Noise Ratio (SNR), defined as follows, \cite{Ellis:2020awk}

\begin{equation}
    \text{SNR} \equiv \sqrt{T \int_{f_{\rm min}}^{f_{\rm max}} \left[\frac{h^2 \,\Omega_{\text{GW}}(f)}{h^2 \,\Omega_{\text{Sens}}(f)} 
    \right]^{2} df}
    \label{SNR_expr}
\end{equation}
where $T$ is the total duration of the experiment. We assume $T$ to be 5 years for all relevant detectors. The quantity $h^2 \Omega_{\text{Sens}}(f)$ represents the experimental sensitivity of a specific configuration to cosmological sources, derived from the power spectral density (PSD) $S_{h}(f)$ \cite{Kuroda:2015owv}.

\begin{equation}
h^2 \,\Omega_{\text{Sens}}(f) = \frac{2 \pi^2}{3 H_{0}^{2}} f^{3} S_{h}(f).
\end{equation}
where $H_{0}=100 \text{h km} \,s^{-1} \text{Mpc}^{-1}$ with $h=0.678\pm 0.009$.

The SNR values for the four benchmark points (BPs) relevant to the detectors are presented in Table \ref{SNRtab}. 
\begin{table}[ht]
    \centering
    \begin{tabular}{|c|c|c|c|}\hline
          Benchmark points&DECIGO-corr&U-DECIGO&U-DECIGO-corr  \\\hline \hline
          BP1&-&-&-\\\hline
          BP4&- &- & $5.49\times10^{6}$\\\hline
          BP5&-&-& $3.88\times10^{6}$\\\hline
          BP6&-&-& $1.42\times10^{6}$\\\hline
    \end{tabular}
    \caption {\it  This table shows the signal-to-noise ratio (SNR) values associated with the final four benchmark points. Dashed lines indicate that the gravitational wave (GW) spectrum corresponding to the benchmark point will not be detectable by the respective detector.}
    \label{SNRtab}
\end{table}
The GW spectrum of BP1 will not be detectable by the upcoming DECIGO GW detectors, while the other three BPs will be detectable by DECIGO. From Table \ref{PTBSM}, we observe that BP4 has the highest order parameter, and thus the highest value of $\alpha$, followed by BP5 and BP6, respectively. Consequently, BP4 exhibits the highest SNR value across all relevant detectors, followed by BP5 and then BP6. DECIGOs are planned to launch around 2030s and will be most sensitive in the frequency band from 0.1 to 10 Hz. They will be operational for 4 to 5 years approximately \cite{Kawamura:2020pcg}.

\section{Conclusion} \label{conclusion}
In this work, we explored the dark matter (DM) phenomenology and the dynamics of strong first-order phase transitions (SFOPT) within a model featuring pseudo-Nambu-Goldstone boson (pNGB) dark matter, realized through a $\mathcal{Z_{\rm 3}}$-symmetric complex scalar, alongside a $\mathcal{Z_{\rm 2}}$-symmetric real scalar both of which are singlet under the $SU(2)_L$ symmetry. Our analysis revealed two primary patterns of strong first-order phase transitions, including transitions along both SM+BSM Higgs direction and BSM Higgs direction only. However, stringent DM direct detection constraints significantly constrain the parameter space, excluding many points where SFOPT was observed. This exclusion underscores the challenge of achieving SFOPT within a viable parameter space, where the complex scalar DM is frequently under-abundant. The dynamics of phase transitions and their associated gravitational wave signatures within a theoretical framework featuring a $\mathcal {Z}_3$-symmetric scalar field have been investigated. By introducing a $\mathcal {Z}_2 $ symmetric real scalar field into this framework, we have identified new parameters— $m_{H_3}, v_\phi$ and $\sin \alpha_2 $ —that significantly affect the strength of the phase transition along the SM Higgs direction.

To explore the implications of these additional parameters, we analyzed seven benchmark scenarios. Three of these scenarios exhibit SFOPT along both the SM and the BSM Higgs field directions, while the remaining four demonstrate SFOPT exclusively along the BSM Higgs direction. The benchmark Point BP1 shows a single-step first-order phase transition with pronounced transitions along all three field directions: $H_1, H_2 $ and $H_3$. The benchmark Points BP2 depicts a two-step first-order phase transitions: the initial step is characterized by a strong transition along the $H_2$ direction, while the subsequent step involves strong transitions along the $H_1 $ field directions. On the other BP3 also exhibits 2 step FOPT similar to BP2, with strong transitions along the $H_1$ (first step) and $H_2$ field direction in the second step. The benchmark Points BP4, BP5, and BP6 all showcase a three-step phase transition, with a strong first-order transition occurring exclusively along the $H_2 $ direction in the initial step. Finally, the benchmark Point BP7 demonstrates a two-step phase transition, characterized by a strong first-order transition along the $H_2$ direction in the first step.

These findings provide valuable insights into the potential gravitational wave signatures associated with phase transitions in this particular theoretical model and contribute to our understanding of the early universe.

Through our analysis, we have shown that part of the model parameters that produce strong first order phase transitions are in direct conflict with the dark matter direct detection constraints provided by the LZ collaboration. For example, the LZ observation rules out the benchmark points BP2, BP3, and BP7, though these benchmark points are consistent with the Xenon 1T limit. The final four benchmark points, that exhibit SFOPT along the SM Higgs field direction show highly underabundant relic density. Even when SFOPT occurs along the BSM Higgs field direction, the resulting scalar DM abundance often falls well below the $2\sigma$ range of the observed relic density. Notably, the direct detection cross-section does not vanish at tree level for zero momentum transfer due to the presence of a single symmetry-breaking scale parameter $\mu_3 $. This parameter is also critical in establishing a barrier in the tree-level potential, enabling first-order phase transitions in the singlet Higgs field direction within a substantial portion of the parameter space while still evading DM direct detection constraints.

Finally, we have checked the testability of the final benchmark points which satisfy both dark matter direct and indirect detection constraints by evaluating the energy density of the gravitational wave spectrum corresponding to those BPs. We have found that the highest peak amplitude of the GW corresponds to BP4, followed by BP5 and BP6 respectively. While the gravitational wave spectrum generated by SFOPT along the SM Higgs direction (BP1) is unlikely to be detectable in near-future experiments, the first-order phase transition signatures along the BSM Higgs direction present a promising avenue for detection in upcoming gravitational wave observatories like DECIGO correlation, U-DECIGO and U-DECIGO correlation.

\vspace{5mm}
\noindent \textbf{Acknowledgments}

The authors would like to thank Abhijit Kumar Saha and Dibyendu Nanda for their collaboration during the early stage of the project. KM would like to thank Sk Jeesun for valuable discussions and acknowledges the financial support provided by the Indian Association for the Cultivation of Science (IACS), Kolkata.

\appendix
\section{Stability bounds on quartic couplings}\label{app:stability}
We have constraints on the quartic coupling constants, first among them is the co-positivity constraint. If there is any matrix $\Lambda$, then the copositivity constraint is given by,
\begin{equation}
    x^T \Lambda x \geq 0
\end{equation}
 for every non-negative vector x.
 In our case, we  have
 \begin{equation}
     x=\begin{bmatrix}
     h^2\\
     
     s^2\\\chi^2\\\phi^2
     \end{bmatrix}
 \end{equation}
 
 \begin{equation}
     \Lambda=\begin{bmatrix}
     \frac{\lambda_{H}}{4}&\frac{\lambda_{HS}}{8}&\frac{\lambda_{HS}}{8}&\frac{\lambda_{H\phi}}{4}\\
     \frac{\lambda_{HS}}{8}&\frac{\lambda_{S}}{4}&\frac{\lambda_{S}}{4}&\frac{\lambda_{S\phi}}{4}\\
    \frac{\lambda_{HS}}{8}&\frac{\lambda_{S}}{4}&\frac{\lambda_{S}}{4}&\frac{\lambda_{S\phi}}{4}\\
    \frac{\lambda_{H\phi}}{4}&\frac{\lambda_{S\phi}}{4}&\frac{\lambda_{S\phi}}{4}&\lambda_{\phi}
      
     \end{bmatrix} =  \begin{bmatrix}
a_{1} & a_{12} & a_{12} & a_{14} \\
. & a_{2} & a_{2} & a_{24} \\
. & . & a_{2} & a_{24} \\
. & . & . & a_{4} 
\end{bmatrix}
 \end{equation}
 
 From the matrix $\Lambda$, we get the co-positivity criteria as,
 \begin{equation}
a_1\geq0 , a_2\geq0, a_4\geq0    
 \end{equation}
 \begin{itemize}
  \item If $ a_{12} \geq0 $, $ a_{14} \geq0 $, $ a_{24} \geq0 $: then for all values. 
  \item  If $ a_{12} \leq 0 $, $ a_{14} \geq0 $, $ a_{24} \geq0 $ $\Rightarrow a_1 a_2 - a_{12}^2 \geq 0$
  \item  If $ a_{12} \geq 0 $, $ a_{14} \leq0 $, $ a_{24} \geq0 $ $\Rightarrow a_1 a_4 - a_{14}^2 \geq 0$ 
  \item  If $ a_{12} \geq 0 $, $ a_{14} \geq0 $, $ a_{24} \leq0 $ $\Rightarrow a_2 a_4 - a_{24}^2 \geq 0$ 
  
  \item If $ a_{12} \geq 0 $, $ a_{14} \leq0 $, $ a_{24} \leq0 $ $\Rightarrow a_4 a_{12} - a_{14} a_{24} + \sqrt{(a_4 a_1 - a_{14}^2)(a_4 a_2 - a_{24}^2)} \geq 0$ 
  
  \item If $ a_{12} \leq 0 $, $ a_{14} \leq0 $, $ a_{24} \geq0 $ $\Rightarrow a_1 a_{24} - a_{12} a_{14} + \sqrt{(a_1 a_2 - a_{12}^2)(a_1 a_4 - a_{14}^2)} \geq 0$ 
  
  \item If $ a_{12} \leq 0 $, $ a_{14} \geq0 $, $ a_{24} \geq0 $ $\Rightarrow a_2 a_{14} - a_{12} a_{24} + \sqrt{(a_2 a_1 - a_{12}^2)(a_2 a_4 - a_{24}^2)} \geq 0$ 
  
  \item If $ a_{12} \leq 0 $, $ a_{14} \leq0 $, $ a_{24} \leq0 $ $\Rightarrow \begin{bmatrix}
   a_1 & a_{12} & a_{14}\\
   a_{12} & a_2 & a_{24}\\
   a_{14} & a_{24} & a_4
  \end{bmatrix}$ must be positive definite
\end{itemize}

\section{Perturbative Unitarity Bounds}
\label{app:PEB}
The two particle neutral states are given by $G^+G^-, \frac{G_0 G_0}{\sqrt{2}},\frac{hh}{\sqrt{2}},\frac{ss}{\sqrt{2}},\frac{\chi\chi}{\sqrt{2}},\frac{\phi\phi}{\sqrt{2}} $ and $h G_0, h s, h\chi, h\phi, G_0s, G_0 \chi, G_0 \phi, s\phi, \chi \phi, s\chi $ and two particle singly charged states are given by $G^+h,G^+\chi,G^+s,G^+G_{0},G^+\phi$. The submatrices required to check perturbative unitarity are as follows:
 \begin{equation}
 \label{pu}
     (\mathcal{M}_{N}^{\text{dir}})_{6\times6}=\begin{bmatrix}
      \lambda_H & \sqrt{2} \lambda_H & \sqrt{2} \lambda_H & \frac{\lambda_{HS}}{\sqrt{2}} & \frac{\lambda_{HS}}{\sqrt{2}} &  \sqrt{2} \lambda_{H\phi} \\
      
      \sqrt{2} \lambda_H & 3 \lambda_H & \lambda_H & \frac{\lambda_{HS}}{2} & \frac{\lambda_{HS}}{2} & \lambda_{H\phi} \\ 
      
      \sqrt{2} \lambda_H & \lambda_H & 3 \lambda_H & \frac{\lambda_{HS}}{2} & \frac{\lambda_{HS}}{2} & \lambda_{H\phi} \\ 
      
      \frac{\lambda_{HS}}{\sqrt{2}} & \frac{\lambda_{HS}}{2} & \frac{\lambda_{HS}}{2} & 3  \lambda_S & \lambda_S & \lambda_{S\phi} \\
      
      \frac{\lambda_{HS}}{\sqrt{2}} & \frac{\lambda_{HS}}{2} & \frac{\lambda_{HS}}{2} &   \lambda_S & 3\lambda_S & \lambda_{S\phi} \\
      
      \sqrt{2} \lambda_{H\phi} & \lambda_{H\phi} & \lambda_{H\phi} & \lambda_{S\phi} & \lambda_{S\phi} & 12 \lambda_{\phi}
     \end{bmatrix}
 \end{equation}
 \begin{equation}
     (\mathcal{M}_{N}^{\text{exc}})_{10\times10}=\rm {diag}(2 \lambda_H, \lambda_{HS}, \lambda_{HS}, 2\lambda_{H\phi}, \lambda_{HS}, \lambda_{HS}, 2\lambda_{H\phi}, 2 \lambda_{S\phi}, 2 \lambda_{S\phi}, 2 \lambda_S )
 \end{equation}
 \begin{equation}
     (\mathcal{M}_{C})_{5\times5}=\rm{diag}(2\lambda_{H}, 2\lambda_{H}, \lambda_{HS},\lambda_{HS},2\lambda_{H\phi})
 \end{equation}
 The distinct eigenvalues are $ 2 \lambda_H, 2\lambda_S, \lambda_{HS}, 2 \lambda_{H\phi}, 2\lambda_{S\phi}, x_{1,2,3,4} $ where $x_{1,2,3,4}$ are the eigenvalues of the submatrix and are the solution of the following Equation,
 \begin{equation}
      x^4 + A x^3 + B x^2 + C x +D  = 0
 \end{equation}
 where,
 \begin{align}
 & A= -12 \lambda_{\phi} - 5 \lambda_H -4 \lambda_S\nonumber\\
 & B=  60 \lambda_{\phi} \lambda_H - 4 \lambda_{H\phi}^2 - 2 \lambda_{HS}^2 + 48 \lambda_{\phi} \lambda_S + 20 \lambda_H \lambda_S - 2\lambda_{S\phi}^2 \nonumber\\
 & C= 2 \lambda_H \lambda_{H\phi}^2 + 24 \lambda_{\phi} \lambda_{HS}^2 + \lambda_H \lambda_{HS}^2 - 240 \lambda_{\phi} \lambda_H \lambda_S + 16 \lambda_{H\phi}^2 \lambda_S - 8 \lambda_{H\phi} \lambda_{HS} \lambda_{S\phi} + 10 \lambda_H \lambda_{S\phi}^2 \nonumber\\
& D= -12 \lambda_{\phi} \lambda_{H} \lambda_{HS}^2 - 8 \lambda_H \lambda_{H\phi}^2 \lambda_S + 4\lambda_H \lambda_{H\phi} \lambda_{HS} \lambda_{S\phi} \nonumber.
    \end{align}

The perturbative unitarity constraints derived from $l=0$ partial wave amplitude such that $|\text{Re}a_0| \leq \frac12$ which translates to $|\text{eig} \mathcal{M}| \leq 8 \pi$: 
\begin{align}
    \lambda_{H} \leq 4\pi,~~ \lambda_{S}\leq 4\pi,~~
     |\lambda_{HS}|\leq 8\pi,~~|\lambda_{H\phi}|\leq 4\pi\,~~
     |\lambda_{S\phi}|\leq 4\pi,~~
     |x_{1,2,3,4}|\leq8\pi.
\end{align}
 
\section{Field dependent mass matrices}\label{sec:FDmassM}
Expressing the Lagrangian in terms of the fields $h$, $s$ and $\phi$, thereafter estimating $M_{ij}=\frac{\partial^2 V}{\partial \phi_i\partial\phi_j}$ where $\phi_i=\{h,s,\phi\}$, we find,
\begin{equation}
\centering
M^2_{\rm scalar} = \begin{bmatrix}
\label{matrix3}
 m_h^2 & h s\lambda_{HS} & 2 h {\phi} \lambda_{H\Phi} & h \chi \lambda_{HS}  \\
    h s \lambda_{HS} & m_s^2 & 2 s {\phi} \lambda_{S\Phi} & 2 s \chi \lambda_{S} - \frac{3\mu_3}{\sqrt{2}}\chi  \\
    2 h {\phi} \lambda_{H\Phi} & 2 s {\phi} \lambda_{S\Phi} & m_{\phi}^2 & 2 \phi \chi \lambda_{S\Phi} \\
    h \chi \lambda_{HS} & 2 s \chi \lambda_{S} - \frac{3\mu_3}{\sqrt{2}}\chi & 2 \phi \chi \lambda_{S\Phi} & m_{\chi}^2\,,
\end{bmatrix}
\end{equation}
where,
\begin{align}
& m_h^2 = 3 \lambda_H h^2+ \frac{\lambda_{HS}}{2} s^2 + \mu_H^2 + \lambda_{H\Phi} \phi^2\,,\nonumber\\  
& m_s^2 = 3 \lambda_S s^2+ \frac{\lambda_{HS}}{2} h^2 + \mu_S^2 + \lambda_{S\Phi} \phi^2 + \lambda_S \chi^2 + \frac{3\mu_3}{\sqrt{2}} s\,,\nonumber\\
& m_{\phi}^2 = \lambda_{H\Phi} h^2 + \lambda_{S\Phi} s^2 + 2 \mu_{\Phi}^2 + 12\lambda_{\Phi} \phi^2 + \lambda_{S\Phi} \chi^2 \,.
\end{align}
The field dependent fermion mass is given by $m_{f_i}=\frac{y_{f_i}}{\sqrt{2}}h$. Similarly, the field-dependent gauge boson masses are expressed as $m_{W_{\pm}}=\frac{1}{2}gh$ and $m_{Z}=\frac{1}{2}\sqrt{g^2+g'^2}h$.  

\section{Counter terms}\label{AppctN}
We intend to ensure that masses of the scalar fields and minima of the scalar fields at zero temperature remain the same as that in the tree level potential even after including the one-loop corrections. These give rise to the following counterterms:
\begin{align}\label{ctexp}
  &    \delta\lambda_{H}=\frac{1}{2v_{h}^3}(\partial_{h}\Delta V - v_{h}\partial^2_{h}\Delta V),\,\,\,
    \delta\lambda_{S}=\frac{1}{2v_{s}^3}(\partial_{s}\Delta V - v_{s}\partial_{s}^2 \Delta V),\,\,\,
    \delta\lambda_{\Phi}=\frac{1}{8v_{\phi}^3}(\partial_{\phi}\Delta V - v_{\phi}\partial_{\phi}^2 \Delta V),\nonumber\\
  & \qquad \qquad \delta\lambda_{HS}=-\frac{\partial_{h}\partial_{s}\Delta V}{v_{h}v_{s}},\,\,\,
    \delta\lambda_{H\Phi}=-\frac{\partial_{h}\partial_{\phi}\Delta V}{2v_{h}v_{\phi}},\,\,\, \delta\lambda_{S\Phi}= -\frac{\partial_{s}\partial_{\phi}\Delta V}{2v_{s}v_{\phi}},\,\,\, \delta\mu_{3}=0,\nonumber\\
  & \qquad \qquad  \delta\mu_{H}^2=\frac{1}{2v_{h}}(-3\partial_{h}\Delta V + v_{s}\partial_{h}\partial_{s}\Delta V +v_{\phi}\partial_{h}\partial_{\phi}\Delta V + v_{h}\partial_{h}^2 \Delta V),\nonumber\\
  & \qquad \qquad \delta\mu_{S}^2=\frac{1}{2v_{s}}(-3\partial_{s}\Delta V + v_{h}\partial_{s}\partial_{h}\Delta V +v_{\phi}\partial_{s}\partial_{\phi}\Delta V + v_{s}\partial_{s}^2 \Delta V),\nonumber\\
  &  \qquad \qquad  \delta\mu_{\Phi}^2=\frac{1}{4v_{\phi}}(-3\partial_{\phi}\Delta V + v_{h}\partial_{\phi}\partial_{h}\Delta V +v_{s}\partial_{\phi}\partial_{s}\Delta V + v_{\phi}\partial_{\phi}^2 \Delta V)\,,
\end{align}
 where $v_{h}=246$\,GeV, $v_{s}$ and $v_{\phi}$ are zero temperature vevs of $h$, $s$ and $\phi$ fields respectively.

\section{Daisy coefficients}\label{sec:DC}
The Daisy coefficients are obtained from finite temperature correction to the effective potential in the high temperature limit ($T\gg m_i$). In the high temperature limit, the Eq. (\ref{finiteT}) takes the form (up to order $T^2$) as
\begin{equation}
\label{vt}
    V_{T} = \frac{T^2}{24}\sum_{i} n_{i}m_{i}^2
\end{equation}
We estimate the Daisy coefficients for the relevant scalar fields from $V_{T\neq 0}^{\rm 1-loop}$ using $\frac{1}{T^2}\frac{\partial^2  V_{T\neq 0}^{\rm 1-loop}}{\partial \phi_i\partial\phi_j}$. We find, 
\begin{align}
   & d_{hh}= \frac{\lambda_{H}}{4}+\frac{\lambda_{HS}}{12}+\frac{\lambda_{H\Phi}}{12}+\frac{3g^2}{16}+\frac{g'^2}{16}+\frac{y^2}{4},\\
   & d_{ss}=\frac{\lambda_{S}}{3}+\frac{\lambda_{HS}}{24}+\frac{\lambda_{S \Phi}}{12},\\
   & d_{\phi\phi} = \frac{\lambda_{H\Phi}}{12}+\frac{\lambda_{S \Phi }}{6}+\lambda_{\Phi},
\end{align}
where $m^2(\phi_i,T)=m^2(\phi_i)+d_i T^2$. For the gauge bosons we obtain,
\begin{equation}
    m_W^2 = \frac{g^2}{4} h^2 + 2g^2 T^2
\end{equation}
\begin{equation}
    m_\gamma^2 = (g^2+g'^2)(T^2+\frac{h^2}{8}) - \frac18\sqrt{(g^2-g'^2)^2(64T^4+16T^2h^2)+(g^2+g'^2)^2h^4}
\end{equation}
\begin{equation}
    m_Z^2 = (g^2+g'^2)(T^2+\frac{h^2}{8}) + \frac18\sqrt{(g^2-g'^2)^2(64T^4+16T^2h^2)+(g^2+g'^2)^2h^4}
\end{equation}

\bibliographystyle{JHEP}
\bibliography{main}
\end{document}